\documentclass[traditabstract]{aa}
\usepackage{graphicx} 
\usepackage{rotating,subfigure,amssymb,afterpage} 
\usepackage{txfonts} 
\usepackage{natbib} 

\newcommand{\propsim}{\lower 3pt \hbox{$\, \buildrel {\textstyle 
      \propto}\over {\textstyle \sim}\,$}} 
\newfont{\gwpfont}{cmssq8 scaled 1000}
\newcommand{\rexcess}{{\gwpfont REXCESS}}
\newcommand{\reflex}{{\gwpfont REFLEX}}

\begin{document} 
   \title{Substructure of the galaxy clusters in the REXCESS sample: 
  observed statistics and comparison to numerical simulations}

   \author{H. B\"ohringer\inst{1}, G.W. Pratt\inst{2,1},  M. Arnaud\inst{2}, S. Borgani\inst{3}, 
    J.H. Croston\inst{4}, T.J. Ponman\inst{5}, S. Ameglio\inst{3,6}, R.F. Temple\inst{5}
   and~ K.~Dolag\inst{7}}
 
   \offprints{H. B\"ohringer, hxb@mpe.mpg.de} 
 
   \institute{$^1$ Max-Planck-Institut f\"ur extraterrestrische Physik, 
                 D 85748 Garching, Germany, {\tt hxb@mpe.mpg.de}\\ 
              $^2$ Laboratoire AIM, IRFU/Service d'Astrophysique - CEA/DSM - CNRS 
                 - Universit\'{e} Paris Diderot, B\^{a}t. 709, CEA-Saclay, F-91191 
                 Gif-sur-Yvette Cedex, France  \\
              $^3$ Dipartimento di Astronomia dell'Universit\'a di
                Trieste, via Tiepolo 11, I-34133 Trieste, Italy\\
              $^4$ School of Physics and Astronomy, University of Southampton, 
                 Southampton, SO17 1BJ, UK \\
              $^5$ School of Physics and Astronomy, University of
                Birmingham, Edgbaston, Birmingham B15 2TT, UK\\
              $^6$ University of Southern California, Los Angeles, CA, USA\\
              $^7$ Max-Planck-Institut f\"ur Astrophysik,
                 D 85748 Garching, Germany\\
             } 
 
   \date{Received xxxx; Accepted yyyy}

\abstract 
{
We study the substructure statistics of a representative sample of galaxy
clusters by means of two currently popular substructure characterisation methods, 
power ratios and centroid shifts. We use the 31 clusters from the \rexcess\ sample, 
compiled from the southern {\it ROSAT} All-Sky cluster survey (\reflex) with a 
morphologically unbiased selection in X-ray luminosity and redshift, all of which 
have been reobserved with {\it XMM-Newton}. The main goals of this work are to study 
the relationship between cluster morphology and other bulk properties, and the 
comparison of the morphology statistics between observations and numerical simulations.
We investigate the uncertainties of the substructure parameters via newly-developed Monte 
Carlo methods, and examine the dependence of the results on projection effects (via the 
viewing angle of simulated clusters), finding that the uncertainties of the parameters 
can be quite substantial. Thus while the quantification of the dynamical state of individual 
clusters with these parameters should be treated with extreme caution, these substructure 
measures provide powerful statistical tools
to characterise trends of properties in large cluster samples. The centre shift 
parameter, $w$, is found to be more sensitive in general and offers a larger 
dynamic range than the power ratios. For the \rexcess\ sample neither the occurence 
of substructure nor the presence of cool cores depends on cluster mass; however a 
weak correlation with X-ray luminosity is present, which is interpreted as selection effect. 
There is a significant anti-correlation between the existence of substantial substructure and 
cool cores. The simulated clusters show on average larger substructure parameters than the 
observed clusters, a trend that is traced to the fact that cool regions are more pronounced 
in the simulated clusters, leading to stronger substructure measures in merging clusters and 
clusters with offset cores. Moreover, the frequency of cool regions is higher in the 
simulations than in the observations, implying that the description of the physical 
processes shaping cluster formation in the simulations requires further improvement.}

 \keywords{X-rays: galaxies: clusters, 
   Galaxies: clusters: Intergalactic medium, Cosmology: observations}  
\authorrunning{B\"ohringer et al.} 
\titlerunning{Substructure in \rexcess\ Galaxy Clusters} 
   \maketitle 
%
 
\section{Introduction} 

There are two major far-reaching interests that motivate our understanding the population of galaxy clusters. Firstly, they are ideal test objects to check the likelihood of a given cosmological model to describe our Universe; secondly,  they are very important probes of
the astrophysical and chemical evolution of the baryonic component of the
Universe (e.g. Rosati et al. 2002, Schuecker et al. 2003a,b, Voit 2005, 
Vikhlinin et al. 2003, 2009, Rozo et al. 2007, Henry et al. 2009, Mantz et al. 2008).  A good understanding and observational characterisation of galaxy clusters is required to attain these goals.

X-ray observations provide an essential window into the study of galaxy 
clusters, as the presence of X-ray radiation implies and traces a well
developed gravitational well (e.g. Sarazin 1986, B\"ohringer 2008), offering the best starting point for the characterisation of the cluster mass and dynamical state. The temperature of the hot intracluster medium is related to the depth of the potential well, and its distribution is related to the dynamical state of the system.
In this study we use statistical measures of substructure observed in the X-ray images, which provide a projected view of the ICM structure, to obtain an impression of the cluster's dynamical state\footnote{Galaxy velocity distributions would provide complementary information on the cluster substructure along the line-of-sight (e.g. Girardi et al. 1997, Biviano et al. 2006).}. Ideally one would like to base such a study on a known relation between a substructure parameterization and a measure of the
deviation from dynamical equilibrium of the cluster. While this could be
an important aim for a future study based on simulations\footnote{Indeed, Yang et al. (2008) have recently made a first approach in this direction by 
comparing substructure measures of simulated clusters with the time since the last merger.}, we pursue here the more qualitative goal of using substructure measures without a calibrated relation  to a physical quantity to sort clusters into a relative ranking order. 

The two direct applications of substructure measures are (i) the study of the influence  
of substructure on cluster scaling relations, e.g., of global X-ray properties, 
and (ii) a comparison of the distribution of the substructure measures in statistically 
representative samples of observed and simulated galaxy clusters.  
\footnote{More than a decade ago one of the applications of determining the ratio of relaxed 
to unrelaxed clusters was aiming at constraining the mean 
matter density of the Universe by means of this parameter 
(e.g. Richstone, Loeb \& Turner 1992, Evrard et al. 1993, Melott, Chambers \&
Miller 2001). Since we have now better approaches for a much more precise determination 
of the cosmic matter density, this application is no more attractive.}
The latter is an important test of how well simulated cluster morphologies correspond to the observed cluster population. Any difference will most probably point to a shortcoming in the description of important physical processes in the simulations. If we want the simulations to correspond well enough to the real world, so that we can draw essential conclusions, we have to perform a series of tests to compare the details of the cluster appearance in simulated and observed objects. While previous studies focussed on reproducing the scaling relations of global cluster parameters (e.g. Borgani et al. 2004, Voit et al. 2003,
McCarthy et al. 2004, 2008, Poole et al. 2006, 2007, 2008, Kay et al. 2007,
Borgani \& Kravtsov 2009), ours  is one of the first such tests to use morphology. 

In the past, a series of different methods have been used to characterise substructure 
in X-ray images of galaxy clusters. Optical substructure has been characterised in 
various ways by e.g. Fitchett \& Webster (1987), Beers et al. 1992 and 
Bird \& Beers (1993). Several of these methods were tested and summarised by 
Pinkey et al. (1996), along with methods that can be applied in a similar way 
to photon distributions. Schuecker et al. (2001) used the methods of Pinkey et al. 
to obtain substructure statistics in X-ray detected galaxy clusters from 
the {\it ROSAT} All-Sky Survey, and tested the correlation of the substructure measures 
with the occurrence of radio halos and with the object's location in the large-scale 
structure environment. Slezak, Durret \& Gerbal (1994) applied a wavelet technique to 
X-ray cluster images.  Gomez et al. (1997, using the Pinkney method), 
Hashimoto et al. (2007b), and Kolokotronis et al. (2001) compared the X-ray and optical 
appearance of  galaxy clusters using ellipticities and center shifts. 
We concentrate here on two methods: centre shifts as a function of the aperture 
radius (e.g. Mohr et al. 1993, 1995; Poole et al. 2006, O'Hara et al. 2006) and 
the determination of so-called power ratios (e.g. Buote \& Tsai 1995, 1996; 
Jeltema et al. 2005, 2008, Valdarini 2006, Ventimiglia et al. 2008). 

Any general characterisation of the galaxy cluster population must be statistical. 
For a comparison study with numerical simulations, a statistically unbiased sample 
of clusters is required that contains a representative distribution of cluster 
morphologies. We therefore base our work on \rexcess\ (Representative {\it XMM-Newton\/} 
Cluster Structure Survey, B\"ohringer et al. 2007), a sample that has been constructed 
with exactly this goal in mind. It is the special construction of the \rexcess\ sample 
that makes a correct comparison with numerical simulations possible, and this is probably 
the most innovative aspect of this paper. Previous publications on the properties of 
the \rexcess\ sample include Pratt et al. (2007, 2009a), on the temperature profiles 
and the X-ray luminosity scaling relations, and Croston et al. (2008), on the gas density 
profiles. Further recent papers describing properties of the clusters in the \rexcess\ 
sample are Pratt et al. (2009b), on the structure of the entropy profiles, 
Arnaud et al. (2009) on the pressure profiles and scaling relations involving the X-ray 
determined equivalent of the Sunyaev-Zeldovich Effect Comptonization
parameter $Y_X$, and a study of the characteristics of the brightest cluster galaxies 
by Haarsma et al. (2009). All of the above papers have made use of the substructure 
parameters obtained by the analysis described herein.

The paper is structured as follows. Section 2 gives a brief description of
\rexcess\ and introduces the substructure characterisation methods used.
Section 3 describes the simulations used and their analysis.
The results of the substructure analysis of the observations
and simulations are discussed in Section 4. Correlations of the
substructure measures with other global clusters parameters are studied
in Section 5 and Section 6 provides a discussion and conclusions.

For the scaling of distance dependent parameters we use a flat concordance 
cosmological model with $H_0 = 70$ km s$^{-1}$ Mpc$^{ -1}$ and $\Omega_m = 0.3$. 

\section{Data Analysis}

\subsection{The \rexcess\ sample}

\rexcess\ was constructed with the aim of providing a statistically well defined collection 
of galaxy clusters, selected only by their X-ray luminosities and redshifts without any bias 
with respect to cluster morphology (for details see B\"ohringer et al. 2007). The full sample
of 33 clusters is drawn from the \reflex\ survey (B\"ohringer et al. 2001, 2004) such as to 
homogeneously cover the luminosity range of 0.4 to 20 $\times 10^{44}\, h_{50}^{-1}$ erg s$^{-1}$ 
in the 0.1 to 2.4 keV band, and the redshifts were chosen so that the clusters fit optimally 
into the field-of-view of the {\it XMM-Newton} X-ray telescopes, with
a cluster free region for background assessment. The resulting redshift range is $z = 0.0564$ 
to 0.1832, with redshift increasing with the luminosity and size
of the clusters. The luminosity range corresponds roughly to an ICM temperature
interval from about 2 to 10 keV (Pratt et al. 2009a). With a requested exposure of at 
least 25 ks, more than 16 ks observation time is left for both EMOS and EPN detectors 
in all but four clusters after removing soft proton flare contamination.

Three of the \rexcess\ targets have two or more well-separated X-ray maxima: the supercluster 
complex Abell 901/902 (RXC\,J0956.4-1004), and the bimodal systems  
RXC\,J2157.4-0747 and RXC\,J2152.2-1942. The latter has a close neighbour detected 
as a well-separated region of X-ray emission in the {\it ROSAT} All-Sky Survey, but 
the two components overlap within their values of $R_{500}$ (defined below) and their 
X-ray emission clearly overlaps in the deeper {\it XMM-Newton} images. The Abell 901/902 
complex contains at least three regions of extended X-ray emission with several additional 
bright X-ray point sources due to AGN. Since the X-ray source regions of these systems 
cannot be treated by the analysis performed in this work without manual decomposition 
which may be subjective, we exclude these two objects from further analysis (they were 
also excluded in Croston et al. 2008 and Pratt et al. 2007, 2009a),
The third system, RXCJ2157.4-0747, has two components that can be well 
separated. In the following analysis the centre is defined to coincide with the centre 
of the main component. The second component then falls inside the aperture corresponding 
to $R_{500}$ (at a distance from the main cluster of $\sim 5 - 7$ arcmin, where $R_{500}$ 
corresponds to 11.1 arcmin) and contributes to the substructure measure\footnote{Note that 
for the determination of the global cluster properties (Pratt et al. 2009a) and density and 
temperature profiles (Croston et al. 2008; Pratt et al. 2007, 2009b) this second component 
was excised.}.

\subsection{Image analysis and scaling}
  
The treatment of the X-ray data and production of the images is described 
in B\"ohringer et al. (2007) and images for all clusters are displayed in Pratt
et al. (2009a). We make use exclusively of the 0.5 to 2.0 keV energy band, since this is approximately the energy range providing X-ray images with the highest signal-to-noise. We combine the data from the three detectors into one composite image which has a minimal number of pixels with zero exposure. Before combining the images, the background is subtracted. The background is obtained from a model fit to a source excised blank sky field, where the model includes homogeneous vignetted and unvignetted
components. Its normalisation is obtained for each detector separately from a comparison to the surface brightness in the outer cluster-free region.
To properly combine the images of the three detectors, we scale the exposure time of the EMOS data sets to the sensitivity ratio between the EPN and EMOS detectors (typically a factor of 3.3). The sensitivity ratio is obtained directly from the data by determining the scaling of the cluster surface
brightness profiles observed with the three detectors. Scaling the exposure map and keeping the summed count image preserves the information on the photon statistics that will become necessary in the course of the analysis. Thus the resulting combined exposure map is in units of effective EPN exposure time.  
Additional X-ray sources in the field-of-view are identified and their extent evaluated using the wavelet-based SAS\footnote{SAS is the ESA provided analysis software system for the reduction of {\it XMM-Newton} observational data. Information can 
be found at: {\tt http://xmm.vilspa.esa.es/sas/}} routine 
{\tt ewavdetect}. The outcome is visually inspected to distinguish significant point sources from extended cluster substructure emission. Non-point sources which correspond to substructure features in the cluster are deleted from
the source list. The sources are then excised from the photon data, with visual checking if the excision radii are sufficient or have to be increased manually. The holes are then filled with the {\it Chandra} CIAO\footnote{CIAO is the publically provided analysis software system for the reduction of {\it Chandra} observational data. Information can be found at: {\tt http://cxc.harvard.edu/ciao/}} routine {\tt dmfilth} using randomisation based on the surface brightness distribution around the holes to fill the gap. The analysis methods described below are then applied to these cosmetically point source-cleaned cluster images. The total area replaced is in all cases a very small fraction of the total image, so that the effect of this cosmetic operation on the photon statistics can be neglected. Excision of the point sources is necessary, however, since their presence near the aperture boundary can severely affect the  results. 

The total number of photons detected per cluster in all three detectors combined ranges from 30\,000 to 170\,000 source photons inside $R_{500}$, except
for two observations with fewer photons that we used as split observations 
(more details in Appendix~\ref{appx:a1}), and for A1689 (RXCJ1311.4-0120), for which more than 300\,000 photons were observed. We therefore have very good photon statistics with which to characterise the cluster morphology. In the following analysis we sample the surface brightness inside a radius of $R_{500}$, which is defined as the radius inside which the mean total mass density  in the cluster is 500 times the critical density of the Universe. The value of $R_{500}$ is not estimated directly from the X-ray derived mass profile, but from the correlation of $M_{500}$ with $Y_X$ (see also Pratt et al. 2009a). The latter parameter, which is the product of the gas mass inside $R_{500}$ and the mean spectroscopic temperature derived from the photons in the radial range $[0.15-0.75]\, R_{500}$, has been found to be a low-scatter mass proxy in numerical simulations (Kravtsov et al. 2006). Its low scatter has been confirmed in the observational analysis of e.g. Arnaud et al. (2007). $M_{500}$ is found by iteration about the relation given by Arnaud et al. (2007), as described by Kravtsov et al. (2006).  $R_{500}$ not only characterises the part of the cluster that is fairly virialised (in non-merging systems) and thus less affected by the matter infall region (e.g. Evrard et al. 1996), but it also marks the region inside which we have a highly significant detection of the X-ray surface brightness for all systems. Therefore this fiducial radius lends itself naturally as the outer limiting radius for our study. 
 
In the following we use two methods for the substructure characterisation. One -- the so-called power ratio method -- is based on a multipole analysis of the azimuthal surface brightness distribution (Buote  \& Tsai 1995, 1996, Buote 2002, Jeltema et al. 2005, 2008, Valdarnini 2006). The other method is based on studying the emission centroid shift as a function of the integration radius (Mohr et al. 1993, 1995, Poole et al. 2006, O'Hara et al. 2006, Kay et al. 2007, Ventimiglia et al. 2008, Yang et al. 2008, Maughan et al. 2008). 

\begin{table*}
\begin{minipage}{\textwidth}
\begin{center}
\caption{Substructure parameters for 31 clusters from the \rexcess\ sample. For this analysis the central cluster emission inside a radius of $0.1 \times R_{500}$ was excised to avoid any influence of cluster cool cores.}
\label{TabR1e}
\centering
\begin{tabular}{l r r r r r r r r r r r }
\hline
\hline
\\
\multicolumn{1}{l}{Cluster} & 
\multicolumn{1}{c}{$P2/P0$} & 
\multicolumn{1}{c}{bias} & \multicolumn{1}{c}{error} & 
\multicolumn{1}{c}{$P3/P0$} & 
\multicolumn{1}{c}{bias} & \multicolumn{1}{c}{error} & 
\multicolumn{1}{c}{$P4/P0$} & 
\multicolumn{1}{c}{bias} & \multicolumn{1}{c}{error} &
\multicolumn{1}{c}{$w$} & \multicolumn{1}{c}{error} \\

\multicolumn{1}{c}{} & 
\multicolumn{1}{c}{$\times 10^{-5}$} & 
\multicolumn{1}{c}{$\times 10^{-5}$} & \multicolumn{1}{c}{$\times 10^{-5}$} & 
\multicolumn{1}{c}{$\times 10^{-7}$} & 
\multicolumn{1}{c}{$\times 10^{-7}$} & \multicolumn{1}{c}{$\times 10^{-7}$} & 
\multicolumn{1}{c}{$\times 10^{-7}$} & 
\multicolumn{1}{c}{$\times 10^{-7}$} & \multicolumn{1}{c}{$\times 10^{-7}$} &
\multicolumn{1}{c}{} & \multicolumn{1}{c}{} \\

(1) & (2) & (3) & (4) & (5) & (6) & (7) & (8) & (9) & (10) & (11) & (12)\\
\\
\hline
\\
{\rm RXCJ0003.8+0203} &   0.205 &   0.0066 &   0.050 &   0.337 &   0.1923 &   0.379 &   0.5507 &   0.083 &   0.334 &   0.0032 &  0.00102\\
{\rm RXCJ0006.0-3443} &   0.546 &   0.0088 &   0.091 &   2.303 &   0.2807 &   1.322 &   0.2767 &   0.127 &   0.281 &   0.0130 &  0.00144\\
{\rm RXCJ0020.7-2542} &   0.135 &   0.0049 &   0.036 &  -0.139 &   0.1777 &   0.372 &   0.3431 &   0.069 &   0.323 &   0.0063 &  0.00078\\
{\rm RXCJ0049.4-2931} &   0.148 &   0.0106 &   0.069 &   0.295 &   0.3261 &   0.851 &   1.8380 &   0.160 &   0.922 &   0.0023 &  0.00078\\
{\rm RXCJ0145.0-5300} &   1.252 &   0.0072 &   0.132 &   1.155 &   0.2146 &   0.787 &   1.9030 &   0.093 &   0.700 &   0.0300 &  0.00141\\
{\rm RXCJ0211.4-4017} &   0.523 &   0.0096 &   0.112 &  -0.223 &   0.3101 &   0.358 &   2.3980 &   0.136 &   0.782 &   0.0046 &  0.00100\\
{\rm RXCJ0225.1-2928} &   0.896 &   0.0238 &   0.205 &   7.652 &   0.7239 &   3.654 &   1.9670 &   0.354 &   1.300 &   0.0121 &  0.00136\\
{\rm RXCJ0345.7-4112} &   0.366 &   0.0120 &   0.073 &   3.363 &   0.3813 &   1.735 &   2.1840 &   0.179 &   0.918 &   0.0052 &  0.00088\\
{\rm RXCJ0547.6-3152} &   0.112 &   0.0028 &   0.021 &   1.822 &   0.0699 &   0.535 &   1.0220 &   0.035 &   0.235 &   0.0070 &  0.00057\\
{\rm RXCJ0605.8-3518} &   0.155 &   0.0026 &   0.027 &   0.009 &   0.0745 &   0.114 &   0.0690 &   0.034 &   0.073 &   0.0059 &  0.00039\\
{\rm RXCJ0616.8-4748} &   0.461 &   0.0079 &   0.110 &   7.580 &   0.2327 &   2.183 &   3.6510 &   0.106 &   0.900 &   0.0131 &  0.00151\\
{\rm RXCJ0645.4-5413} &   0.416 &   0.0036 &   0.059 &  -0.110 &   0.1084 &   0.159 &   0.2678 &   0.051 &   0.181 &   0.0039 &  0.00042\\
{\rm RXCJ0821.8+0112} &   0.207 &   0.0253 &   0.109 &   9.254 &   0.8668 &   4.260 &   2.4590 &   0.372 &   1.617 &   0.0045 &  0.00144\\
{\rm RXCJ0958.3-1103} &   0.211 &   0.0063 &   0.048 &   0.038 &   0.1830 &   0.357 &   0.2421 &   0.086 &   0.203 &   0.0034 &  0.00072\\
{\rm RXCJ1044.5-0704} &   0.321 &   0.0021 &   0.031 &  -0.035 &   0.0613 &   0.055 &  -0.0092 &   0.027 &   0.051 &   0.0072 &  0.00040\\
{\rm RXCJ1141.4-1216} &   0.098 &   0.0033 &   0.023 &   1.129 &   0.0917 &   0.457 &   0.1163 &   0.040 &   0.136 &   0.0054 &  0.00059\\
{\rm RXCJ1236.7-3354} &   0.038 &   0.0086 &   0.030 &   0.227 &   0.2701 &   0.620 &   0.2089 &   0.117 &   0.418 &   0.0052 &  0.00069\\
{\rm RXCJ1302.8-0230} &   1.233 &   0.0071 &   0.129 &   3.304 &   0.2152 &   1.100 &   1.3020 &   0.094 &   0.553 &   0.0153 &  0.00086\\
{\rm RXCJ1311.4-0120} &   0.035 &   0.0009 &   0.007 &   0.041 &   0.0206 &   0.051 &   0.0121 &   0.009 &   0.020 &   0.0040 &  0.00029\\
{\rm RXCJ1516.3+0005} &   0.099 &   0.0035 &   0.023 &   0.415 &   0.1030 &   0.328 &   0.3115 &   0.043 &   0.187 &   0.0037 &  0.00044\\
{\rm RXCJ1516.5-0056} &   0.636 &   0.0075 &   0.091 &  10.250 &   0.2445 &   2.058 &   1.0280 &   0.106 &   0.479 &   0.0176 &  0.00140\\
{\rm RXCJ2014.8-2430} &   0.073 &   0.0020 &   0.015 &   0.614 &   0.0597 &   0.300 &   0.1227 &   0.026 &   0.093 &   0.0058 &  0.00030\\
{\rm RXCJ2023.0-2056} &   0.056 &   0.0155 &   0.051 &  -0.338 &   0.5603 &   0.610 &   0.8529 &   0.252 &   0.809 &   0.0167 &  0.00147\\
{\rm RXCJ2048.1-1750} &   0.840 &   0.0051 &   0.080 &   3.638 &   0.1479 &   0.866 &   2.0140 &   0.066 &   0.456 &   0.0419 &  0.00432\\
{\rm RXCJ2129.8-5048} &   0.037 &   0.0079 &   0.090 &   2.966 &   0.2349 &   1.220 &  -0.0584 &   0.107 &   0.269 &   0.0419 &  0.01997\\
{\rm RXCJ2149.1-3041} &   0.011 &   0.0046 &   0.012 &   1.951 &   0.1294 &   0.685 &   0.0450 &   0.059 &   0.082 &   0.0034 &  0.00053\\
{\rm RXCJ2157.4-0747} &   1.501 &   0.0266 &   0.275 &  -0.847 &   0.9897 &   0.954 &   3.4010 &   0.444 &   2.032 &   0.1080 &  0.90430\\
{\rm RXCJ2217.7-3543} &   0.061 &   0.0037 &   0.022 &   0.705 &   0.1116 &   0.364 &   0.2425 &   0.046 &   0.203 &   0.0018 &  0.00373\\
{\rm RXCJ2218.6-3853} &   0.762 &   0.0022 &   0.052 &   0.514 &   0.0728 &   0.287 &   0.1893 &   0.028 &   0.099 &   0.0155 &  0.00049\\
{\rm RXCJ2234.5-3744} &   0.144 &   0.0040 &   0.027 &  -0.085 &   0.1166 &   0.181 &   0.3276 &   0.055 &   0.148 &   0.0075 &  0.00056\\
{\rm RXCJ2319.6-7313} &   0.694 &   0.0089 &   0.138 &  -0.198 &   0.3012 &   0.799 &   0.5018 &   0.126 &   0.346 &   0.0217 &  0.00085\\

\\
\hline
\end{tabular}
\end{center}

NOTES: The power ratio parameters have been determined for an aperture with a radius 
of $R_{500}$. The corresponding results without center excision are given in Table A.1 
in the Appendix. For each of the power ratio parameters we provide the value of the 
noise contribution to the power ratio result (bias) which has been subtracted from 
the measured result to provide the value listed in columns 2, 5, and 8. The uncertainties 
determined from the Poissonisation simulations are listed in columns 4, 7, and 10 (error). 
The center shift statistic parameter $w$ and its uncertainty are listed in columns 11 and 12.

\end{minipage}
\end{table*}
%
%


\subsection{Power ratios}

The power ratio method introduced by Buote \& Tsai (1995) is motivated by the
idea of identifying the X-ray surface brightness as a representation of the projected 
mass distribution of the cluster. The power ratio is then a multipole decomposition of 
the potential of the two-dimensional, projected mass distribution. Following the recipe 
of Buote \& Tsai, the moments, $P_n$ are determined as follows

\begin{equation}
P_0 = \left[ a_0 \ln (R_{ap}) \right]^2
\end{equation}

\begin{equation}
P_m = { 1 \over 2 m^2 R_{ap}^{2m} } \left( a_m^2 + b_m^2 \right)
\end{equation}

\noindent where $R_{ap}$ is the aperture radius. The moments $a_m$ and $b_m$ are calculated using:

\begin{equation}
a_m(R) = \int_{R^\prime \le R_{ap}} S(x^\prime) (R^\prime)^m \cos (m\phi^\prime)~
d^2x^\prime
\end{equation}

\noindent and

\begin{equation}
b_m(R) = \int_{R^\prime \le R_{ap}} S(x^\prime) (R^\prime)^m \sin (m\phi^\prime)~ d^2x^\prime,
\end{equation}

\noindent where $ S(x)$ is the X-ray surface brightness, $x$ is used as a label for the pixel, 
and the integral extends over all pixels inside the aperture radius. $a_0$ in Eqn.~1 is thus 
the total radiation intensity inside the aperture radius.

Since all terms $P_m$ are proportional to the total intensity of the
cluster X-ray emission, while only the relative contribution of the higher
moments to the total emission is of interest, the multipole expansion
power terms are normalised by $P_0$, resulting in the so-called power ratios, $P_m/P_0$. 
Similarly to all previous studies, we only make use of the lowest moments from 
$P_1$ to $P_4$ (the dipole, quadrupole, hexapole and octopole). The dipole, $P_1$, 
is used to find the centre of symmetry of the X-ray surface brightness distribution. 
We use a minimisation routine (the simplex method of Press et al. 1992) to find the 
minimum of $P_1$ as a function of input centre coordinates. The disappearance of  
the dipole indicates that the signal is well balanced in opposite  directions from 
the centre at all position angles, and we take this as a sign that the cluster is 
correctly centred (with respect to the radial weighting scheme used to calculate 
$P_1$). We checked visually that this defines the centre of symmetry, and most 
often also the central maximum of the cluster's X-ray emission, and found the 
method to be reliable. We then used this centre, defined for a vanishing dipole 
signal, to determine the power ratios from $P_2/P_0$ to $P_4/P_0$. $P_2/P_0$ describes 
the quadrupole of the surface brightness  distribution and is not necessarily a measure 
of substructure; a quadrupole will also be detected for a very regular elliptical cluster, 
which could be well relaxed. In practice, low to moderate values of $P_2/P_0$ are found for 
regular elliptical clusters, while larger values of $P_2/P_0$ are a sign of cluster mergers. 
The lowest power ratio moment providing a clear substructure measure is thus $P_3/P_0$. 
$P_4/P_0$ describes substructure on slightly finer scales and is found to be correlated 
to $P_2/P_0$ in this and previous studies.

The results from the power ratio analysis  determined within $R_{500}$ are given in 
Table~1 for core-excised cluster images and in Table~A.1 for the full aperture.  
In Table~2 we also give the cluster centres in sky coordinates, as determined from 
the dipole  minimisation for the innermost aperture with no center excision. The Tables 
also give two measures of the uncertainty on the power ratios, as we discuss below and 
in more detail in the Appendix.

To assess whether we have detected a significant deviation from azimuthal symmetry, we must 
consider the effect of photon noise. This noise first introduces a bias since even a 
completely symmetric cluster would be detected with some residual structure due to 
the photon noise affecting real observations. We estimate this bias from Monte Carlo simulations of
azimuthally randomised cluster images, and subtract this signal from the
power ratios. The statistical uncertainties of significant substructure signal, again due 
to photon noise, are then estimated from a second set of
Monte Carlo simulations in which Poisson noise is added to the observational 
data and uncertainties are determined from standard deviations. These uncertainties are 
larger than the bias for clusters with significant substructure signal. These methods 
of error estimation are further explained in the Appendix, where we also quantify how 
realistic these estimated uncertainties are.    

To explore the effects of cool cores in the clusters, we have undertaken the power ratio 
analysis both with and without excising the central region ($r \le 0.1\, R_{500}$). From 
the above power ratio definition it is obvious that for large values of $R_{\rm ap}$ the 
exclusion of the central region has little influence on the magnitude of the derived power 
ratios. The question of whether the central regions are retained or excised is more important 
for the second method of structure assessment described in the next Section.

\subsection{Centre shifts}

The centre shift method is based on a measurement of the variation in separation between the X-ray peak and the centroid calculated within an increasing aperture size. For our analysis we use two centre definitions: the centres found in the $P_1$ minimisation procedure described above, and centres found by the determination of a local maximum in smoothed images\footnote{We have also used manually determined local maxima and the maximum obtained from minimisation of $P_1$ in the innermost region ($r < 0.1\,R_{500}$). The results we obtained from the different approaches are quantitatively similar and demonstrate the robustness of the centre shift parameters obtained.}. The latter is determined from the X-ray surface brightness peak on an image smoothed with a Gaussian with $\sigma$ of $\sim 8$ arcsec (in images where point sources have been removed). We determine the centre shift between the local maximum and the centroids obtained by $P_1$ minimisation within 10 apertures ($r \le n \times 0.1\, R_{500}$, with n = 1,2 ..10). For the runs where the central region ($r \le 0.1 R_{500}$) is excised, we have centroids within 9 apertures. The fiducial $w$ parameter is then the  standard deviation of the different centre shifts (in units of $R_{500}$), defined as (see also Poole et al. 2006):

\begin{equation}
w~ =~  \left[ {1 \over N-1}~ \sum \left( \Delta_i -  \langle \Delta \rangle \right)^2 \right] ^{1/2} 
~\times~ {1 \over R_{500} } 
\end{equation}

\noindent where $\Delta_i$ is the distance between the X-ray peak and the centroid of the $i$th aperture.

The uncertainties in the centre shifts and in the $w$ parameter are determined 
with the same simulations as the uncertainties of the power ratios, i.e., by using Poissonised resampled cluster X-ray images. The standard deviation of the $w$ parameter in the simulation results is used as an estimate of the measurement uncertainties. We do not subtract a noise bias for the $w$-parameter as in the case of the power ratios. The results of the centre shift analysis are
given for core-excised cluster images in Table~1  and for the full aperture in Table~A.1. Table~2 gives the cluster centres from the determination of  the local maxima, in sky coordinates, as used in the centre shift analysis. 

\section{Simulations}

The representative nature of \rexcess, and the homogeneous nature of the associated X-ray observations, makes it an ideal sample for comparison with numerical simulations. 
What interests us most here is the question of whether the simulated clusters resemble the clusters
observed in our Universe, when the comparison is based on a representative sample of real clusters.

The set of simulated clusters that we use for this comparison comprises
117 clusters identified from the hydrodynamical simulation of a large
cosmological volume in a $\Lambda$CDM model, presented by Borgani et
al. (2004). These clusters have virial masses in the range $0.8 \times
10^{14}$ to $1.3 \times 10^{15}\,h^{-1}$ M$_{\odot}$.  As such, the
sample covers a range of ICM emission weighted temperatures, $T_{\rm ew}$,
between 1.18 and 7.1 keV. Since there is only one cluster
with a temperature larger than 5 keV, we supplemented the sample with 
four massive objects taken from the set of simulated clusters presented by Dolag et al. (2008). These additional systems have emission--weighted temperatures in the range $T_{\rm ew} = 8.1 - 12.6$ keV and virial masses in the range $1.0$ to $2.2
\times 10^{15}\,h^{-1}$ M$_{\odot}$. The mass and ICM temperature 
distribution of the simulated clusters is thus different 
from that of the \rexcess\ sample and we discuss this in terms of a fair comparison below. All simulations have been carried
out with the TreePM-SPH GADGET-2 code (Springel 2005). They included
the treatment of radiative cooling, a uniform time-dependent UV
background, and a sub-resolution model for star formation and energy
feedback from galactic winds (see Springel \& Hernquist 2003 for
details). All clusters are identified from the simulations at $z=0$.

   \begin{figure*}
\begin{center}
  \includegraphics[height=6cm]{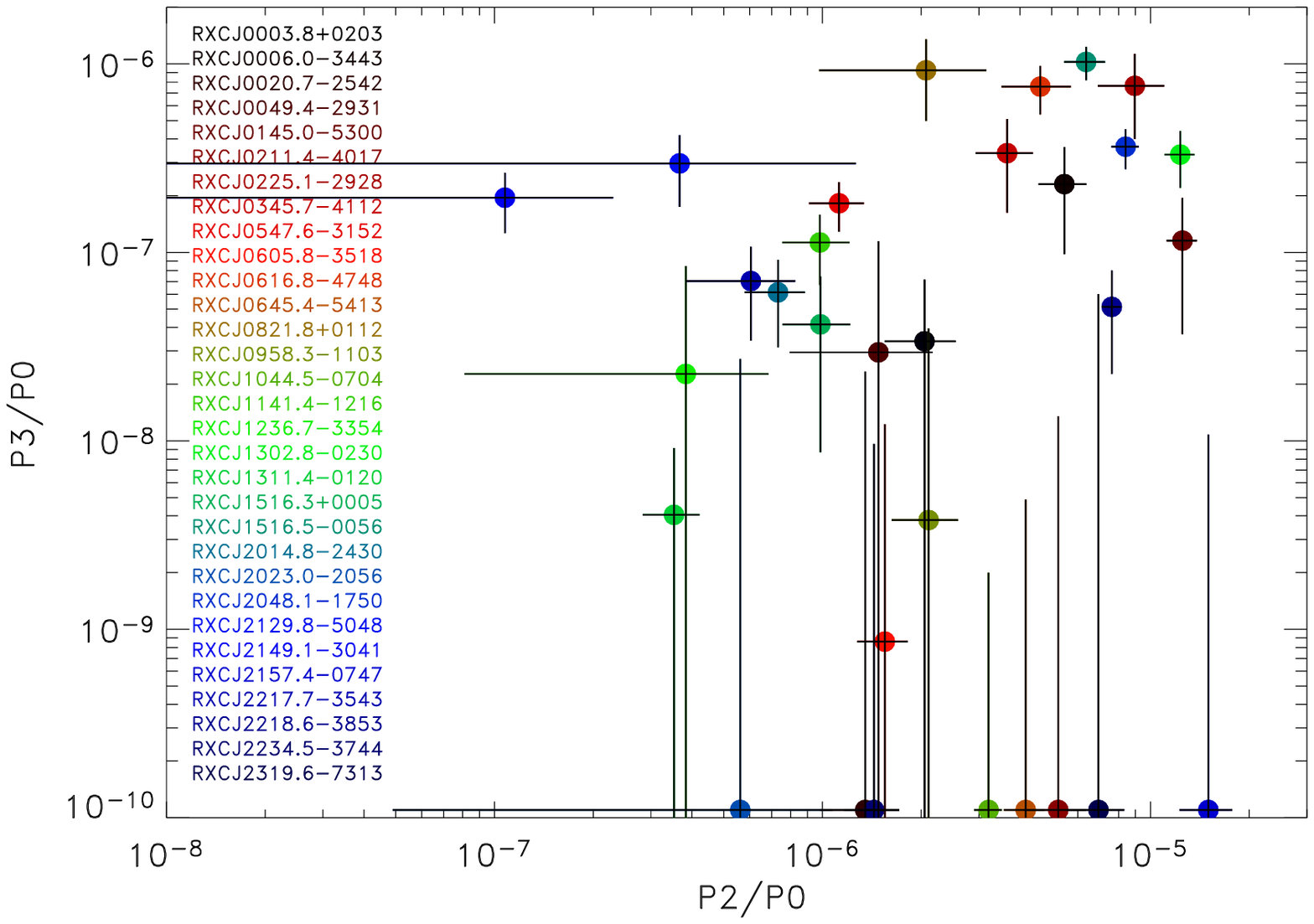}
\hfill
  \includegraphics[height=6cm]{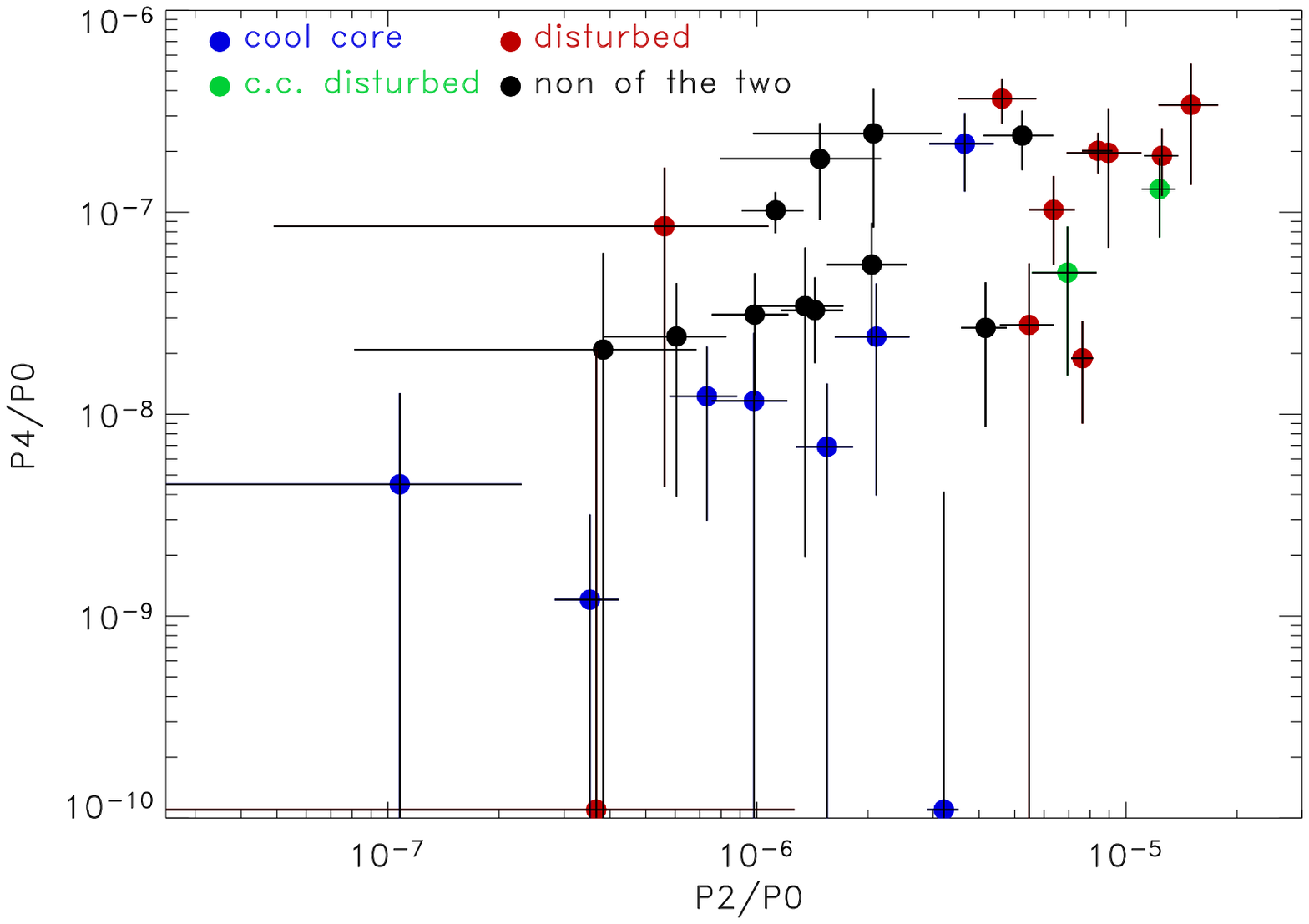}
  \includegraphics[height=6cm]{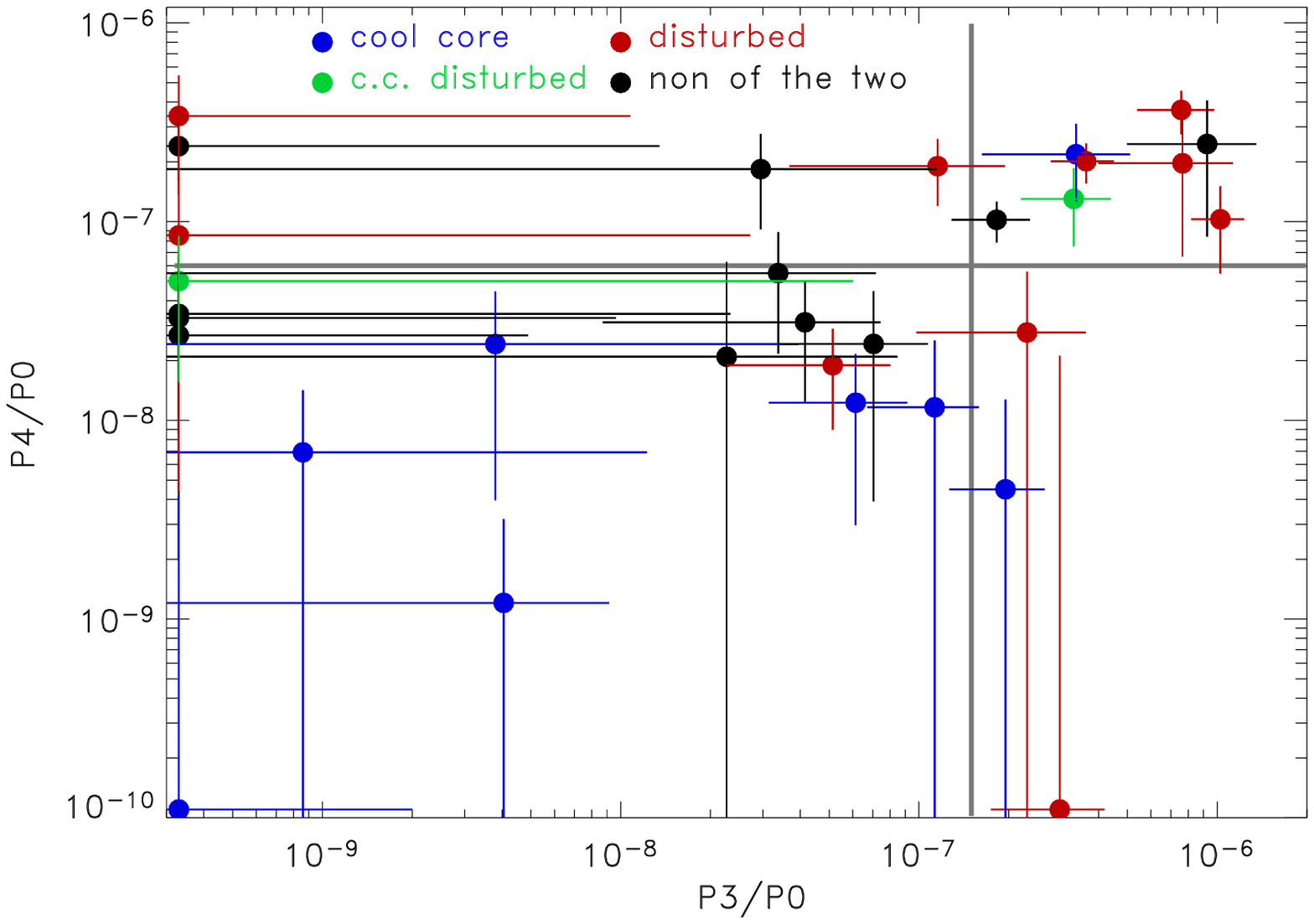}
\hfill
  \includegraphics[height=6cm]{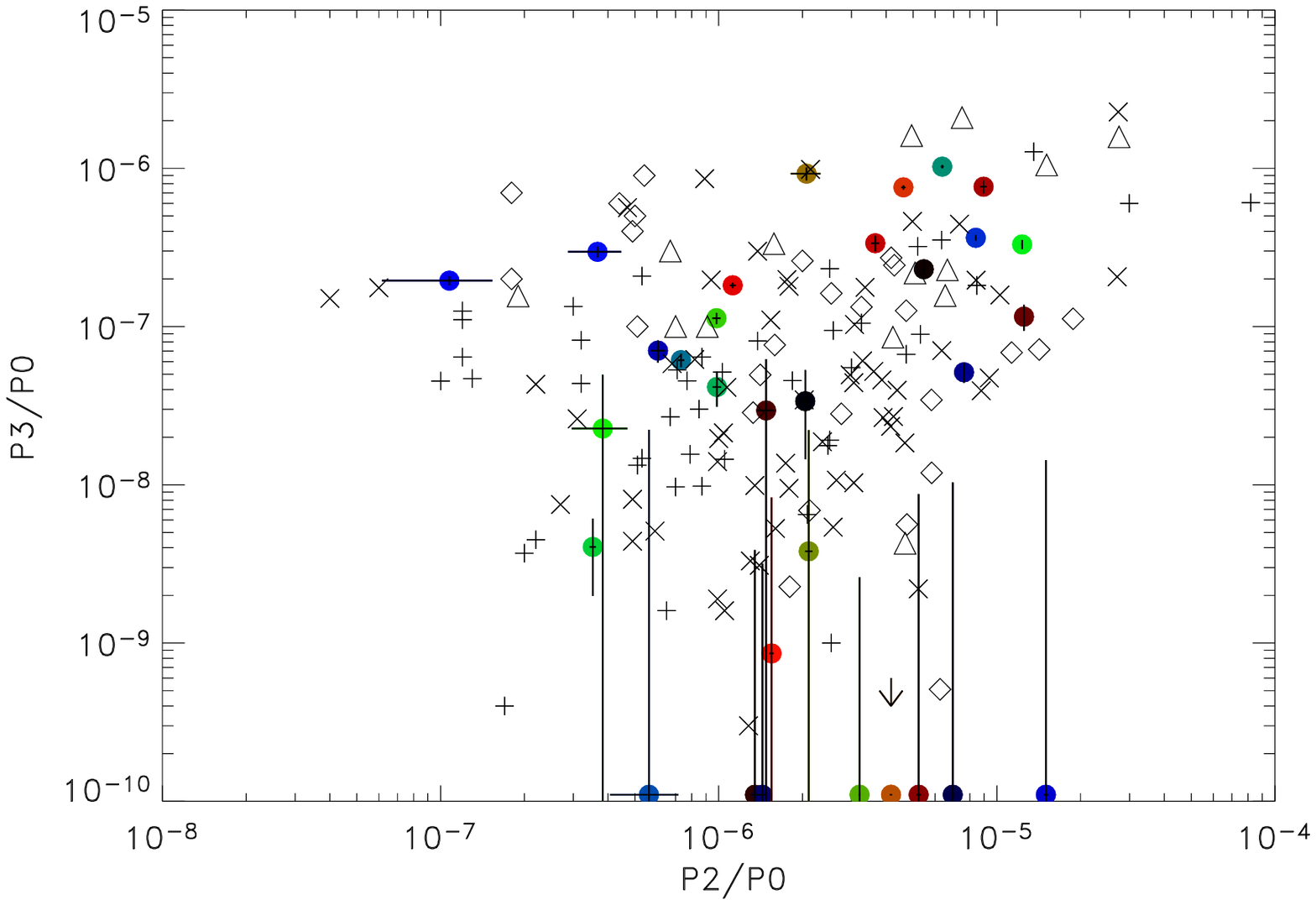}    
\end{center} 
      \caption{Results of the power ratio analysis for the 31 clusters of the \rexcess\ sample, determined for photons the radial range [0.1 - 1] $R_{500}$. The  {\it upper left, upper right} and {\it lower left} panels show the distribution of $P2/P0$, $P3/P0$, and $P4/P0$ in pairs. The uncertainties shown are determined by Monte Carlo simulations with Poisson noise added to the image pixels. While in the upper left panel we identify the clusters by their name in colors, we use some cluster properties for the identification of the data points for the upper right and lower left panel (as used in Pratt et al. 2009a, the threshold for a cluster to be considered as morphologically disturbed is a $\langle w \rangle$ parameter $\ge 0.01$). In the lower left panel we define a tentative classification for a cluster to be morphologically disturbed in terms of the $P3/P0$ and $P4/P0$ parameters, shown by the grey lines (numerical values given in the text). The {\it lower right} panel shows a comparison of our P2/P0 and P3/P0 parameter results and those from the literature. The diamonds and triangles are data from Jeltema et al. (2005) for clusters with $z < 0.45$ and $z > 0.45$, respectively. The $\times$ and $+$ symbols correspond to the data from Buote \& Tsai (1996) for power ratios determined with apertures $R_{ap} \le 0.7 h_{80}^{-1}$ Mpc and $R_{ap} > 0.7 h_{80}^{-1}$ Mpc, respectively. In this plot the uncertainties on the \rexcess\ data are calculated from the azimuthal randomization test (see Fig. A.1, and accompanying text).}
         \label{Fig1}
   \end{figure*}

X-ray images for these clusters were produced and analysed by Ameglio et al. (2007,
2009). They have also been used for a similar sub-structure analysis
in connection with the mass-temperature relation of clusters
by Ventimiglia et al. (2008). X--ray emission within each pixel has been
computed by accounting for the contribution of all the gas particles
falling in projection within that pixel, as described by Ameglio et
al. (2007).  The images used here have been obtained for the energy
range 0.5 to 2 keV, which is the same energy band used for the
analysis of {\it XMM-Newton} observations. The synthetic X-ray images do not
include any X-ray background and photon noise, thus the outcome of the analysis 
of the simulated images is interpreted as results without statistical uncertainties. 
For all aspects of the analysis carried out in this paper, this makes the interpretation 
easier and has no effect on the conclusions from the
comparison with real data. The centre of each map of
X--ray emissivity coincides in projection with the minimum of the
cluster gravitational potential. Each image has a size of 256 by 256
pixels and covers a region of $4 \times R_{500}$ around each
cluster. The simulated cluster images have a
resolution of $\sim 0.031\, R_{500}$, comparing well to the
resolution of observations, which are characterised by a typical half
energy width of $0.030 - 0.033\, R_{500}$. As we discuss in the
Appendix, the resolution at which the X--ray emission of real clusters
is observed has no significant effect on the results of the
substructure analysis at the range of resolution powers relevant here.
Any difference in the resolution between simulated and \rexcess\ 
cluster maps is therefore not important for the
purpose of our analysis. 


\section{Results of the morphological analysis}

In this Section we first present the observational results, then discuss some implications of the analysis of the simulations, and finally, we compare the observations to the simulations.

\subsection{Power ratios} 

The first three panels in Fig.~1 show the results of the power ratio analysis
in pairs of two of the power ratio parameters. As detailed above, values shown are those
for an aperture of $R_{500}$ (with the central $0.1\,R_{500}$ excised) so as to  
provide a global characterization of the clusters. While for the upper left and 
lower right panel the clusters are colour coded to identify them by their name, 
we characterise the clusters in the other two panels by their properties (cool cores, 
having central densities $h_{70}^{-2}~ n_{e0} > 4 \times 10^{-2}$ cm$^{-3}$, and 
morphologically disturbed clusters, having a $w$ parameter $> 0.01$)
\footnote{These classifications were used Pratt et al. (2009a) to isolate the most 
extreme thirty per cent of the sample according to those criteria, and are further 
discussed in Sect.~\ref{sec:censhift}.}. The values of $P_2/P_0$ (i.e., the quadrupole moments) 
are larger than the higher order ratios. Even  though the data have very good photon statistics 
compared to the average X-ray observation, the results have substantial photon noise 
uncertainties, implying that the power ratio method becomes insufficiently sensitive 
for images of much lower quality. The $P_2/P_0$ (quadrupole) versus $P_4/P_0$ (octople) 
plot shows, as in previous studies, the strongest correlation.  However, the plot of 
$P_3/P_0$ vs. $P_4/P_0$ also shows a correlation, such that both parameters will most 
likely flag the same cluster as either regular or disturbed. With respect to the cool 
core and morphological disturbance criteria mentioned above,  we note that in both 
the $P_2/P_0-P_4/P_0$ pair and the $P_3/P_0-P_4/P_0$ pair most clusters designated 
as disturbed accumulate in the upper right quadrant and most cool core clusters 
accumulate in the lower left quadrant.


We now try to set a tentative threshold criterion for the dynamical state of a system 
in terms of the power ratio parameters $P_3/P_0$ and $P_4/P_0$. We find that a value 
of $P_3/P_0 = 1.5 \times 10^{-7}$ separates out 11 non-symmetric clusters (about one 
third of the sample), and a value of $P_4/P_0 = 6 \times 10^{-8}$ separates out 13 clusters. 
These threshold criteria are shown as horizontal and vertical grey lines in the lower 
left panel of Fig.~1.

The three clusters with no significant $P_3/P_0$ but high $P_4/P_0$ are, from top to bottom in the plot: RXCJ2157.4-0747, RXCJ0211.4-4017, and RXCJ2023.0-2056. RXCJ2157.4-0747 is the double component cluster which features a large centre shift and large quadrupole and octopole moments, but just lacks a hexapole moment. RXCJ0221.4-4017 appears rather regular but obtains a significant octopole moment due to some distortion in the outer surface brightness contour. The third cluster in this group, RXCJ2023.0-2056 is azimuthally symmetric on large scale but has an Eastern extension of the central region. 

Similarly we find three clusters separated out by the $P_3/P_0$ parameter as disturbed but having a low $P_4/P_0$ value: RXCJ0006.0-3443, RXCJ2149.1-3041, and RXCJ2129.8-5048. The first cluster 
has no significant octopole moment, the second cluster is fairly regular but displays a significant hexapole moment due to some distortions in the outer surface brightness contour, and RXCJ2129.8-5048 has a low surface brightness elongation to the NE but no significant octopole moment.
One may thus consider a combination of $P_3/P_0$
and $P_4/P_0$ measures, with the threshold criteria defined above, as a good general characterisation of significant substructure in a given system. However, distortions near the aperture radius can have a strong influence on the derived power ratios.


\subsection{Centre shifts}
\label{sec:censhift}

\begin{table}
\begin{center}
\caption{Coordinates (J2000) of the cluster centers for 31 objects of the \rexcess\ cluster sample.}
\label{TabR1e_cen}
\centering
{\tiny
\begin{tabular}{l r r r r r}
\hline
\hline
\\
\multicolumn{1}{l}{Cluster} & 
\multicolumn{2}{c}{Local minimum} & 
\multicolumn{2}{c}{$P_1$ minimum} & \multicolumn{1}{c}{F} \\
\cline{2-5}

\multicolumn{1}{c}{} & 
\multicolumn{1}{c}{RA } & 
\multicolumn{1}{c}{DEC } & 
\multicolumn{1}{c}{RA } & 
\multicolumn{1}{c}{DEC } & 
\multicolumn{1}{c}{} \\
\\
\hline
\\
{\rm R0003+0203}& 00~03~49.7&  02~03~57.5 & 00~03~49.6&  02~03~51.9& 1  \\
{\rm R0006-3443}& 00~05~59.9& -34~43~23.0 & 00~06~00.4& -34~43~19.3& 1  \\
{\rm R0020-2542}& 00~20~42.2& -25~42~24.5 & 00~20~41.9& -25~42~36.0& 1  \\
{\rm R0049-2931}& 00~49~23.0& -29~31~13.9 & 00~49~23.0& -29~31~11.8& 1  \\
{\rm R0145-5300}& 01~44~59.7& -53~01~03.1 & 01~44~58.1& -53~01~12.9& 1  \\
{\rm R0211-4017}& 02~11~24.8& -40~17~28.3 & 02~11~25.1& -40~17~28.4& 1  \\
{\rm R0225-2928}& 02~25~09.3& -29~28~36.3 & 02~25~08.5& -29~28~36.4& 1  \\
{\rm R0345-4112}& 03~45~46.2& -41~12~13.8 & 03~45~46.4& -41~12~15.7& 1  \\
{\rm R0547-3152}& 05~47~38.4& -31~52~12.2 & 05~47~38.5& -31~52~10.9& 1  \\
{\rm R0605-3518}& 06~05~54.2& -35~18~08.7 & 06~05~54.0& -35~18~08.3& 1  \\
{\rm R0616-4748}& 06~16~51.7& -47~47~40.4 & 06~16~51.5& -47~47~42.2& 1  \\
{\rm R0645-5413}& 06~45~29.3& -54~13~40.4 & 06~45~29.4& -54~13~38.9& 1  \\
{\rm R0821+0112}& 08~21~50.9&  01~11~52.4 & 08~21~50.7&  01~11~56.1& 1  \\
{\rm R0958-1103}& 09~58~22.3& -11~03~53.6 & 09~58~22.1& -11~03~50.3& 1  \\
{\rm R1044-0704}& 10~44~33.0& -07~04~08.6 & 10~44~32.9& -07~04~07.7& 1  \\
{\rm R1141-1216}& 11~41~24.4& -12~16~37.4 & 11~41~24.4& -12~16~38.8& 1  \\
{\rm R1236-3354}& 12~36~41.3& -33~55~37.2 & 12~36~41.0& -33~55~26.3& 1  \\
{\rm R1302-0230}& 13~02~53.3& -02~31~00.4 & 13~02~52.7& -02~30~56.9& 1  \\
{\rm R1311-0120}& 13~11~29.5& -01~20~27.7 & 13~11~29.4& -01~20~30.1& 1  \\
{\rm R1516+0005}& 15~16~18.1&  00~05~27.8 & 15~16~18.0&  00~05~23.1& 1  \\
{\rm R1516-0056}& 15~16~44.2& -00~58~11.9 & 15~16~44.5& -00~58~21.8& 1  \\
{\rm R2014-2430}& 20~14~51.7& -24~30~19.9 & 20~14~51.7& -24~30~21.6& 1  \\
{\rm R2023-2056}& 20~22~58.8& -20~56~56.1 & 20~22~59.0& -20~56~57.2& 1  \\
{\rm R2048-1750}& 20~48~12.2& -17~51~19.8 & 20~48~10.4& -17~50~35.1& 0  \\
{\rm R2129-5048}& 21~29~36.5& -50~48~52.2 & 21~29~40.9& -50~48~54.8& 0  \\
{\rm R2149-3041}& 21~49~07.6& -30~42~04.7 & 21~49~07.6& -30~42~04.6& 1  \\
{\rm R2157-0747}& --        & --          & 21~57~29.5& -07~47~54.9& 0  \\
{\rm R2217-3543}& 22~17~45.5& -35~43~30.1 & 22~17~45.6& -35~43~31.0& 1  \\
{\rm R2218-3853}& 22~18~40.3& -38~54~05.7 & 22~18~40.1& -38~54~00.6& 1  \\
{\rm R2234-3744}& 22~34~28.0& -37~43~52.1 & 22~34~27.1& -37~44~02.3& 0  \\
{\rm R2319-7313}& 23~19~40.2& -73~13~38.4 & 23~19~39.9& -73~13~36.1& 1  \\

\\
\hline
\end{tabular}
}
\end{center}

NOTES: The first set of coordinates designate the local X-ray maximum found in X-ray 
images (0.5 - 2 keV band) smoothed with a Gaussian with $\sigma \sim 8$ arcsec. 
The second set of coordinates is derived from minimising the dipole ($P_1$) signal in 
the determination of the power ratios in the smallest aperture with a radius of 
$0.1\times R_{500}$. The flag, F, indicates which cluster centre has been used 
for the determination of the centre shift parameter $w$. For flag 1 the local 
maximum of the smoothed images was used (default), for clusters with flag 0 the P1 
minimisation result was used.
\end{table}



The $w$-parameters found for the \rexcess\ sample are given in Table~1, and the centre coordinates used for this analysis are compared to those obtained from the dipole minimisation in Table~2. The two centres are generally very similar.
Exceptions are the three clusters with very diffuse, low surface brightness regions in the centre, RXCJ2048.1-1750, RXCJ2129.8-5048, and RXCJ2157.4-0747. Here the local maximum is difficult to define automatically, even with smoothing, and the minimisation of the dipole moment marks a much more reliable central location as determined by visual inspection of the images. While in the first two cases the offset is moderate, for RXCJ2157.4-0747 the centre determination is not stable, and we do not give results for the local maximum for this cluster. Therefore we recommend minimisation of the dipole moment to determine the centre for these and similar systems, as indicated by the flag in Table~2. In addition, for RXCJ2234.5-3744, which shows some substructure in the central region, the dipole moment centring provides a better indication of the local maximum.     

\subsection{Comparison with previous work}

In the lower right panel of Fig. 1 we compare the range of our results for the $P_2/P_0$ versus $P_3/P_0$ correlation to the previous observational results from  Buote \& Tsai (1996) and Jeltema et al. (2005). We note that the 
parameter range covered is very similar for all the studies. Some of
the data points from the work of Jeltema et al. extend to somewhat higher values in both parameters, as indicated by the points in the upper right corner. However, for these points no photon noise bias was subtracted,
as it was for our data, and some of the clusters in the Jeltema et al. sample have large photon noise and would possibly have significantly reduced values if the photon noise correction were to be applied. Therefore we conclude that the parameter range covered by observational studies up to date is very similar, even though the selection of the cluster sample is not strictly equivalent.

Maughan et al. (2008) have studied a large sample of 115 galaxy clusters in the redshift range 0.1 to 1.3 observed with Chandra, and analysed the cluster morphologies with the same centre shift parameter technique as the one used in the present work. They find a very similar distribution of $w$ parameter values, lying in the range 0.0007 - 0.0695.

\subsection{Comparison of power ratios and centre shifts}

   \begin{figure}
   \begin{center}
   \includegraphics[width=\columnwidth]{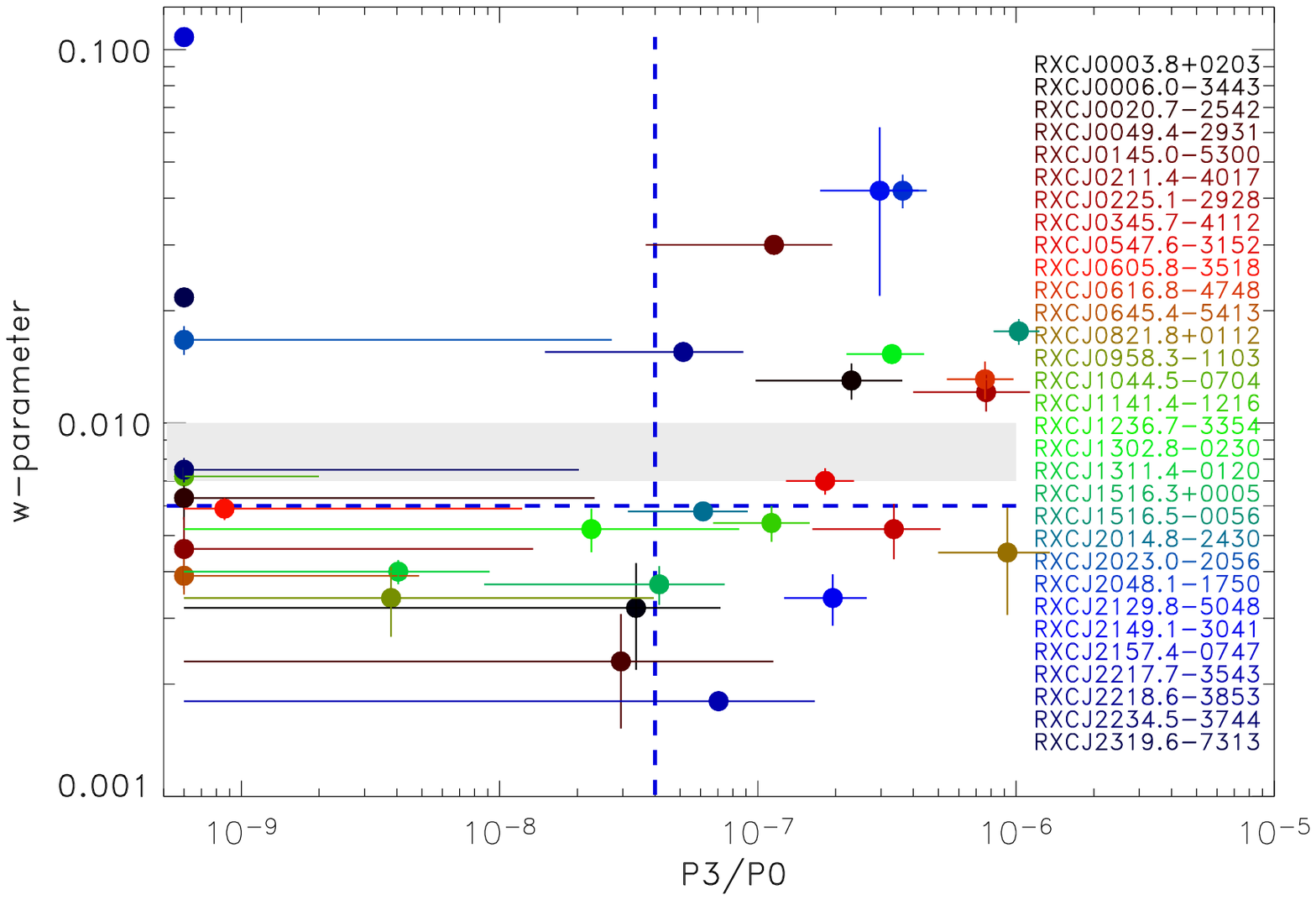}
   \includegraphics[width=\columnwidth]{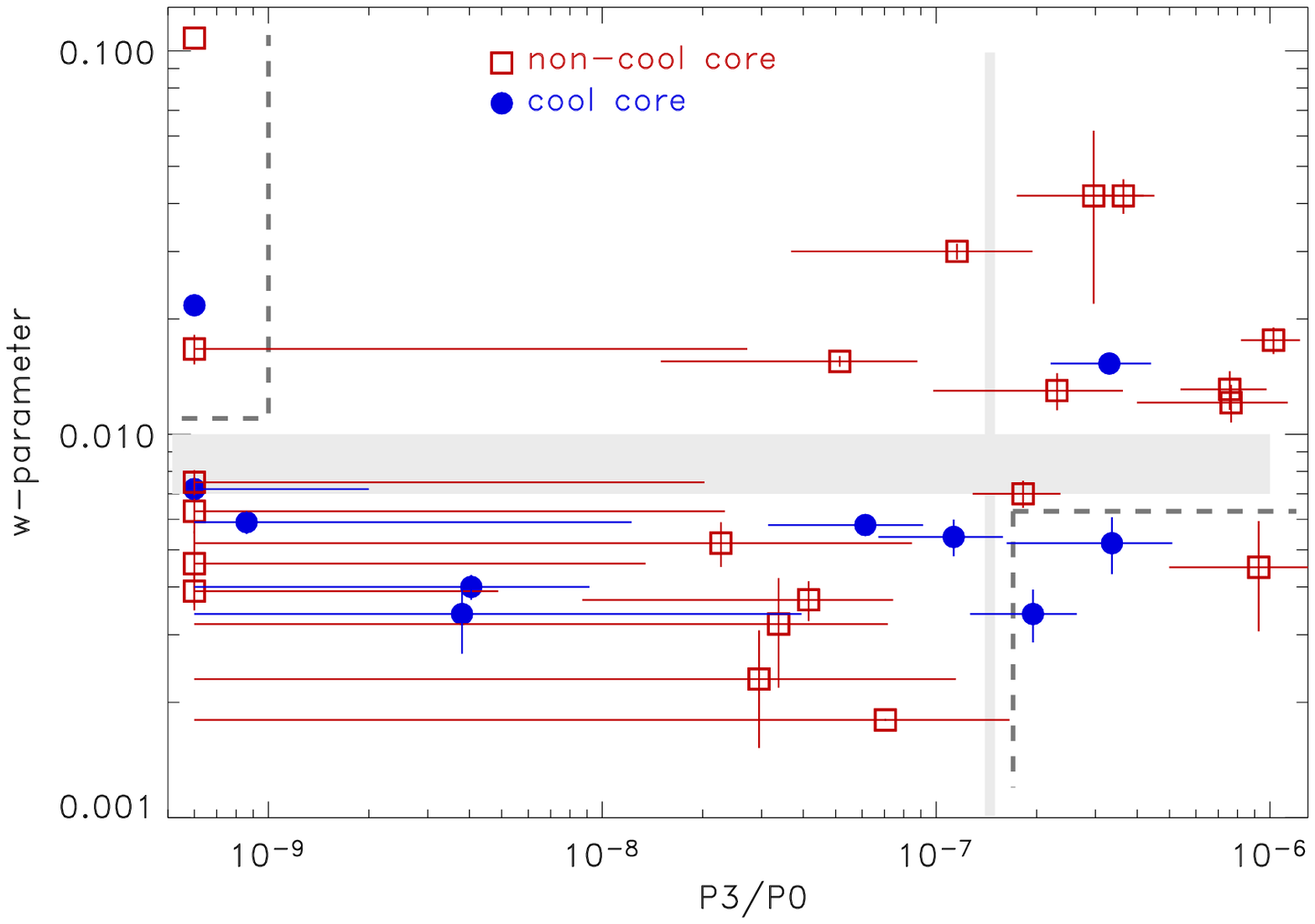}
      \caption{Correlation of the $w$ parameter with $P3/P0$ for the 31 clusters of 
the \rexcess\ sample, with both parameters derived from core excised images. 
{\it Upper panel:\/} Clusters identified by name. The dashed blue lines divide the 
sample roughly in two halves with respect to both parameters. {\it Lower panel:\/} 
Cool core clusters identified. The thin grey line marks the tentative $P3/P0$ threshold 
criterion for morphological disturbance as defined in Fig. 1. The dashed grey lines 
enclose the clusters with discrepant power ratio vs $w$ classification, as further 
discussed in the text. In both panels the grey stripe marks the gap in the $w$-parameter 
distribution used for further discussion of the distribution statistics in the text.}
         \label{Fig2}
  \end{center}
   \end{figure}

\subsubsection{Sorting clusters}

The clusters marked in red in the lower left hand panel of Fig.~1 were identified as the thirty per cent of the sample that were most disturbed according to the $w$ parameter (see Pratt et al. (2009a) and Section 4.2). Of these, only one cluster, RXCJ2218.6-3853, has no large hexapole or octopole moment; however, this system is elongated with a significant quadrupole moment and has a large isophotal centre shift in the central region.
In addition, two cool core clusters, RXCJ1302.8-0230 (green) and RXCJ0345.7-4112 (blue) are among the clusters classified as disturbed through the $P_3/P_0$ vs. $P_4/P_0$ classification. While RXCJ0345.7-4112 shows a very regular central region, it has a very low surface brightness extension in its Eastern outer regions; RXCJ1302.8-0230 has a cool core but with a clear offset from the overall cluster symmetry.

In Fig.~2 we plot $w$ versus $P_3/P_0$ for the 31 clusters of the \rexcess\ 
sample\footnote{For the results plotted in the Figure, the central region was excised, but 
including the central region gives very similar results (cf. corresponding Table~\ref{TabR1e_1} 
in the Appendix).}. We have selected the $P_3/P_0$ parameter for this comparison since it is the 
lowest multipole moment that gives an unambiguous signature of dynamical distortion (since an 
elliptical cluster can be relaxed). While there is a clear correlation between the two parameters, 
there is also a large scatter, illustrating that the two methods are weighing structural features 
in different ways. The horizontal and vertical lines in the upper panel 
($P_3/P_0 = 4 \times 10^{-8}$; $w = 0.006$) divide the sample roughly in half with 
respect to the two parameters, with fifteen objects in the  lower (left) half and sixteen in 
the upper (right) half of the Figure. Alternatively, using the larger gap in the $w$-parameter 
distribution  ($w = 0.007 - 0.01$, marked by a grey stripe in the Figure), we find that 
21 clusters are found in the lower left and upper right quadrants, while only 10 clusters 
fall near the boundaries into the other two quadrants, which implies that with these dividing 
lines 68\% of the clusters would be classified in the same way and 32\% differently using 
these parameter cuts. 

Inspecting the clusters which have discrepant substructure classifications in terms
of $P_3/P_0$ and $w$ parameter (Fig.~2), we easily find the reason for the discrepancy. 
The three clusters in the upper left of Fig.~2, having no significant value of $P_3/P_0$, 
are: RXCJ2157.4-0747 (at the top) which is the
double component cluster which features a large centre shift and large quadrupole and 
octopole moments, but lacks a hexapole moment. Similarly, RXCJ2319.6-7313 is elliptical 
in large scale morphology, with a bright, elongated central region, resulting in a strong 
centre shift and dipole moment but vanishing hexapole moment.
RXCJ2023.0-2056, is azimuthally symmetric on large scale but has an Eastern 
extension of the central region. Among the outliers above the $P_3/P_0$-threshold in 
the lower right corner of Fig.~2, RXCJ0821.8+0112 is regular in the central region 
except for a substructure clump near $R_{500}$, RXCJ0345.7-4112 shows a very regular 
central region, but has an very low surface brightness extension in the Eastern outer 
region, and  RXCJ2149.1-3041 is quite regular but distorted near  $R_{500}$ as discussed above.

In the lower panel of Fig.~2 we designate the clusters by their cool core properties. 
While only two cool core clusters, RXCJ1302.8-0230 and RXCJ2319.6-7313, are classified 
by the $w$ parameter as disturbed, several cool core clusters are found to have high 
$P_3/P_0$ values.  The two cool core clusters with the largest values of $P_3/P_0$ 
are RXCJ0345.7-4112 and RXCJ1302.8-0230 which have already been discussed above. 
As already noted, the difference in the classification obviously stems from the 
stronger weighting that the power ratios give to the outer regions. The $P_3/P_0$ 
parameter is very sensitive to substructure in the outskirts, while the centre shift 
parameter is very sensitive to isophotal structure in the inner region. This result 
is further explored in Section A.3 in the Appendix, where we study power ratios with 
varying aperture radius.  

\subsubsection{Sensitivity}

The uncertainties derived from the Monte Carlo simulations for the $w$-parameter are in most cases smaller than the corresponding uncertainties for $P_3/P_0$. In particular, a useful signal/noise value is still obtained for small values of $w$, in contrast to the results for $P_3/P_0$. The log-mean relative errors for $P_2$ are 22 per cent (excluding 1 cluster with a result consistent with zero) and for $P_3$ they are 70 per cent (excluding the 8 clusters with 
signal less than zero). For the $w$ parameter the log-mean error is 15 per cent  for all clusters. To evaluate the sensitivity of the method the uncertainties should be considered in the context of the dynamic range of the parameter values. Since for both $P_3/P_0$ and $w$ the values span a range of about two orders of magnitude ($10^{-8} - 10^{-6}$ for $P_3/P_0$ and $0.001 - 0.1$ for $w$), the relative errors can directly be compared. Figure~3 illustrates the significance of both methods: the typical uncertainty for $P_3/P_0$ is comparable to the parameter value, while for $w$ it is more of the order of 10 per cent. Such a comparison clearly shows that $w$ appears to be more sensitive than $P_3/P_0$. 

   \begin{figure}
   \begin{center}
   \includegraphics[width=\columnwidth]{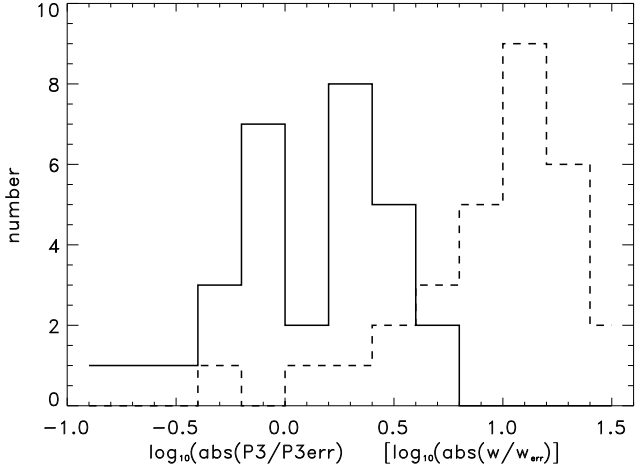}
      \caption{Signal-to-noise ($P3/\Delta(P3)$ ) and ($w/\Delta w$) distribution of the  
    power ratios (solid line) and centre shifts (dashed line)  for the 31 \rexcess\ clusters 
    (plotted in logarithmic units).}
         \label{Fig3}
  \end{center}
   \end{figure}

The $w$ parameter also gives a better overall measure of the deviation from symmetry 
and is not discriminating against 
the central regions as do the power ratios (as illustrated by the examples discussed above). Hence we decided that the $w$ parameter better suits the goal of substructure characterisation in the \rexcess\ project, and we have made use  of this substructure measure in our other \rexcess\ papers. Our practical approach to separating the disturbed clusters from the regular objects takes advantage of the observed gap in the $w$ parameter distribution around $w = 0.01$. We designate clusters above this limit as disturbed, leaving 12 clusters with the classification of being dynamically distorted - about one third of the sample. 

\begin{table}
\begin{center}
\caption{Correlation tests of various substructure measures using the 31 galaxy 
    clusters in the \rexcess\ sample.}
\label{Tab4}
\centering
\begin{tabular}{l r r r r r}
\hline
\hline
\\
\multicolumn{1}{l}{Correlation} & 
\multicolumn{1}{c}{$\tau$} & 
\multicolumn{1}{c}{$P$} & \multicolumn{1}{c}{$\rho$} &
\multicolumn{1}{c}{$P$} & \multicolumn{1}{c}{Figure} \\
(1)       &     (2)        &   (3)         &    (4)          &  (5)       & (6)  \\
\hline
\\
$P_3/P_0-w\ (R_{500})$     & 0.80 & 0.43 & 0.21 & 0.25 & 2\\
$P_3/P_0-w\ (R_{500})^a$   & 1.22 & 0.22 & 0.29 & 0.12 & 2\\
$P_3/P_0-w\ (0.9R_{500})$  & 1.63 & 0.10 & 0.29 & 0.10 & -\\
$P_3/P_0-w\ (0.8R_{500})$  & 2.20 & 0.03 & 0.37 & 0.044 & -\\
$P_3/P_0-w\ (0.7R_{500})$  & 2.13 & 0.03 & 0.38 & 0.035 & A.5\\
$P_3/P_0-w\ (0.6R_{500})$  & 1.38 & 0.17 & 0.25 & 0.17 & -\\
$P_3/P_0-w\ (0.5R_{500})$  & 0.36 & 0.72 & 0.01 & 0.99 & -\\
\\
\hline
\\
$M_{500}-w$    & 0.24 & 0.81 &-0.04 & 0.81 & 11 \\
$M_{500}-P_3/P_0$ & 0.39 & 0.69 &-0.06 & 0.73 & 11 \\
$L_1-w$       & 0.49 & 0.63 & -0.09 & 0.63 & 12\\
$L_1-P3/P0$   & 1.48 & 0.14 & -0.22 & 0.22 & 12\\
$M_{500}-L_{\rm rat}$    & 0.03 & 0.97 & -0.04 & 0.84 & 14\\
$L_1-L_{\rm rat}^b$ & 1.41 & 0.16 & 0.245 & 0.18 & 14\\
$L_2-L_{\rm rat}$ & 0.43 & 0.67 & 0.055 & 0.76 & 14\\
$n_{e0} - P3/P0$         & 1.54 & 0.12 & -0.30 & 0.10 & 13\\
$t_{\rm cool} - P3/P0$       & 1.86 & 0.06 &  0.34 & 0.06 & 13\\
$n_{e0} - w$                & 2.57 & 0.01 & -0.52 & 0.004 & -\\
$t_{cool}- w$                & 2.62 & 0.009 & 0.50 & 0.006 & -\\
\\
\hline
\end{tabular}
\end{center}

NOTES: Column (1) lists the correlation tested, (2) gives the parameter of Kendall's $\tau$ test and (3) the corresponding probability of a null-correlation, (4) gives the result for Spearman's rank test  $\rho$ and (5) the corresponding probability and (6) gives the Figure number that shows the correlation. For the correlation analysis the ASURV software package (Isobe et al. 1986) was used. $^a$ same correlation as in the first row but excluding the outlier object RXCJ2157.4-0747. $^b$ $L_1$ is the X-ray luminosity in the $R_{500}$ aperture in the [0.1-2.4] keV band and $L_2$ refers to the core excised luminosity (values given in Pratt et al. 2009a).
\end{table}
%
%


To quantify the correlation of the two substructure measures we analyse the correlation statistics of the data by means of the Kendall and Spearman tests. To estimate the test statistics we use the analysis  package ASURV (Astronomical Survival Statistics, Isobe et al. 1986), which tests for correlations in the presence of censored data. Table~3 lists the results of the correlation of the $w$ vs. $P_3/P_0$ parameter, which gives probabilities of 0.43 and 0.25 for no correlation according to Kandell's $\tau$ and Spearman's $\rho$ rank test, respectively, indicating a weak correlation. The correlation improves significantly if we remove one outlier, RXCJ2157.4-0747 (the two component cluster, visible in the top left in both panels of Fig.~2). In this case the corresponding probability for no correlation decrease to 0.22 and 0.12, giving stronger significance to the correlation. We also studied how the correlation changes for power ratios determined with smaller apertures and found a significant improvement in the correlation statistics for aperture sizes of $0.7 - 0.8\, R_{500}$, as listed in Table~3 and as further discussed in the Appendix.

   \begin{figure}
   \begin{center}
   \includegraphics[width=\columnwidth]{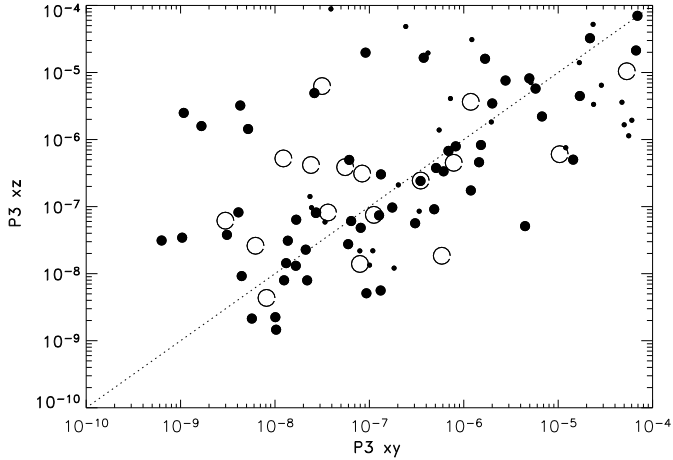}
      \caption{Comparison of the $P_3/P_0$ result for two different orthogonal projections of the simulated clusters, showing a clear correlation with a large scatter. The larger points mark clusters with temperatures above 2 keV, the open symbols those with $T_X$ above 3.5 keV. The dashed line indicates equality of both parameters.}
         \label{Fig4}
  \end{center}
   \end{figure}

   \begin{figure}
   \begin{center}
   \includegraphics[width=\columnwidth]{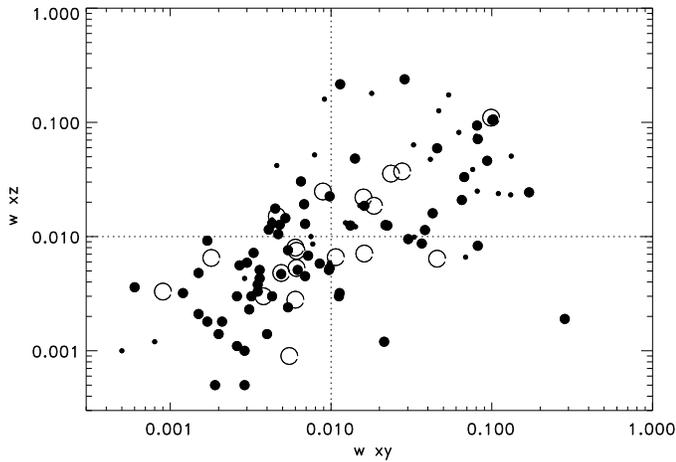}
      \caption{Comparison of the centroid shift statistic parameter, $w$, for two different orthogonal projections of the simulated clusters, showing a clear correlation with a large scatter. The larger dots mark clusters with temperatures above 2 keV, the open symbols those with $T_X$ above 3.5 keV. The dashed lines indicate the defined threshold for considering a cluster as dynamically distorted for both viewing directions.}
         \label{Fig5}
  \end{center}
   \end{figure}

\section{Observations versus simulations} 

\subsection{Dependence on viewing angle in simulations}

The simulations provide us with the means for another important approach to test
the significance of the methods for the characterisation of substructure. Since for the 121 simulated clusters, images from three different orthogonal viewing angles are at hand, we can test how much the substructure characterisation varies depending on viewing angle. Therefore we can straightforwardly investigate how well the structure parameter results are correlated for the three different projections of each cluster. Figure~4 shows the results for two projections of the power ratio $P_3/P_0$. There is a clear correlation and also a very large scatter.

Similarly we compare the correlation of the w-parameter determined for different viewing angles in Fig.~5. The correlation as shown in the plot seems tighter than that for $P_3/P_0$. We have determined the mean orthogonal scatter (standard deviation from the diagonal) of the power ratios and the $w$ parameter for all three projection pairs, averaging using both linear and logarithmic summation. The results are summarised in Table~4. The standard deviation of the values is about as large as the values themselves; however, we note that the mean orthogonal scatter for $P_3/P_0$ is about twice as large as the scatter for the center shift parameter, $w$.

{\footnotesize
\begin{table}
\begin{center}
\caption{Comparison of the mean orthogonal scatter of the substructure parameters from the analysis of the simulated clusters seen from different projection angles. \label{Tab3}}

\centering
\begin{tabular}{l r r r r }
\hline
\hline
\\
\multicolumn{1}{l}{Parameter pair} & 
\multicolumn{1}{c}{X-Y} & \multicolumn{1}{c}{X-Z} & 
\multicolumn{1}{c}{Y-Z} & \multicolumn{1}{c}{mean} \\
\\
\hline
\\
$\langle \Delta \widetilde{P_2}/\widetilde{P_2} \rangle /\sqrt{2}$    & 0.72 & 0.79 & 0.77 & 0.76 \\
${\rm exp}\ {\langle ln( \Delta \widetilde{P_2}/\widetilde{P_2}) \rangle} /\sqrt{2}$ & 0.48 & 0.56 & 0.57 & 0.54 \\ 
$\langle \Delta \widetilde{P_3}/\widetilde{P_3} \rangle /\sqrt{2}$    & 0.99 & 1.85 & 0.91 & 1.25 \\
${\rm exp}\ {\langle ln( \Delta \widetilde{P_3}/\widetilde{P_3}) \rangle} /\sqrt{2}$ & 0.63 & 0.83 & 0.72 & 0.73 \\ 
$\langle \Delta \widetilde{P_4}/\widetilde{P_4} \rangle  /\sqrt{2}$       & 1.13  & 1.03 & 1.01 & 1.06 \\
${\rm exp}\ {\langle  ln(\Delta \widetilde{P_4}/\widetilde{P_4}) \rangle} /\sqrt{2}$ & 0.71 & 0.80 & 0.80 & 0.77 \\
$\langle \Delta w/w  \rangle /\sqrt{2}$           & 0.50 & 0.49 & 0.48 & 0.49 \\
${\rm exp}\ {\langle (ln(\Delta w/w)) \rangle} /\sqrt{2}$   & 0.32 & 0.29 & 0.26 & 0.29 \\
\\
\hline
\end{tabular}
\end{center}

NOTES: $\widetilde{P_2} = P_2/P_0$, etc. The orthogonal scatter is defined as the mean deviation from the diagonal in the plot and thus the algebraic expressions in the Table contain an extra factor of $1/\sqrt{2}$. The mean is determined by  both linear and logarithmic averaging. The means for the three projections and the total mean are given. The standard deviation of these parameters from the mean is slightly smaller than the means, but of the same order of magnitude. 

\end{table}

%
%
%

}

   \begin{figure}
   \begin{center}
   \includegraphics[width=\columnwidth]{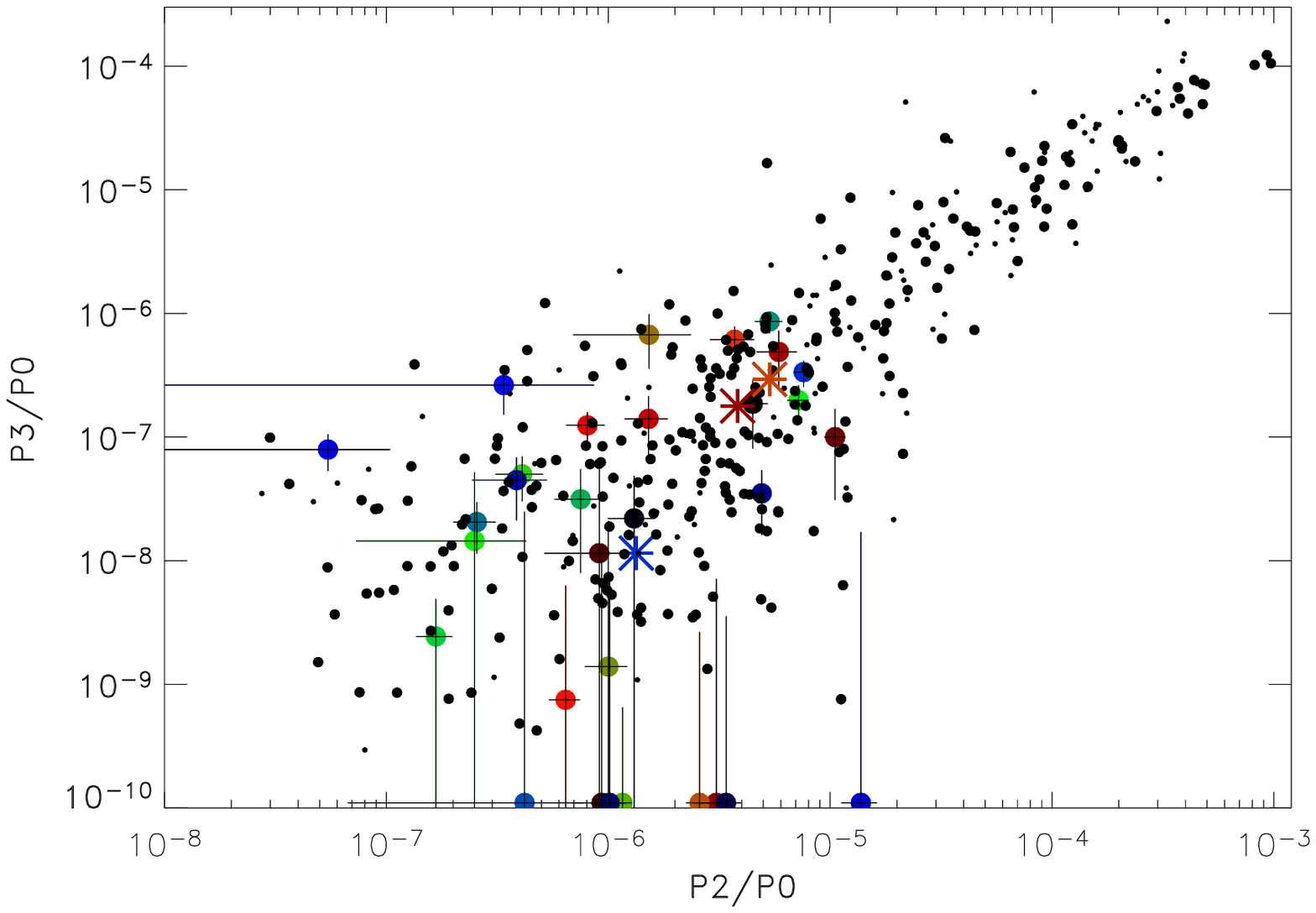}
   \includegraphics[width=\columnwidth]{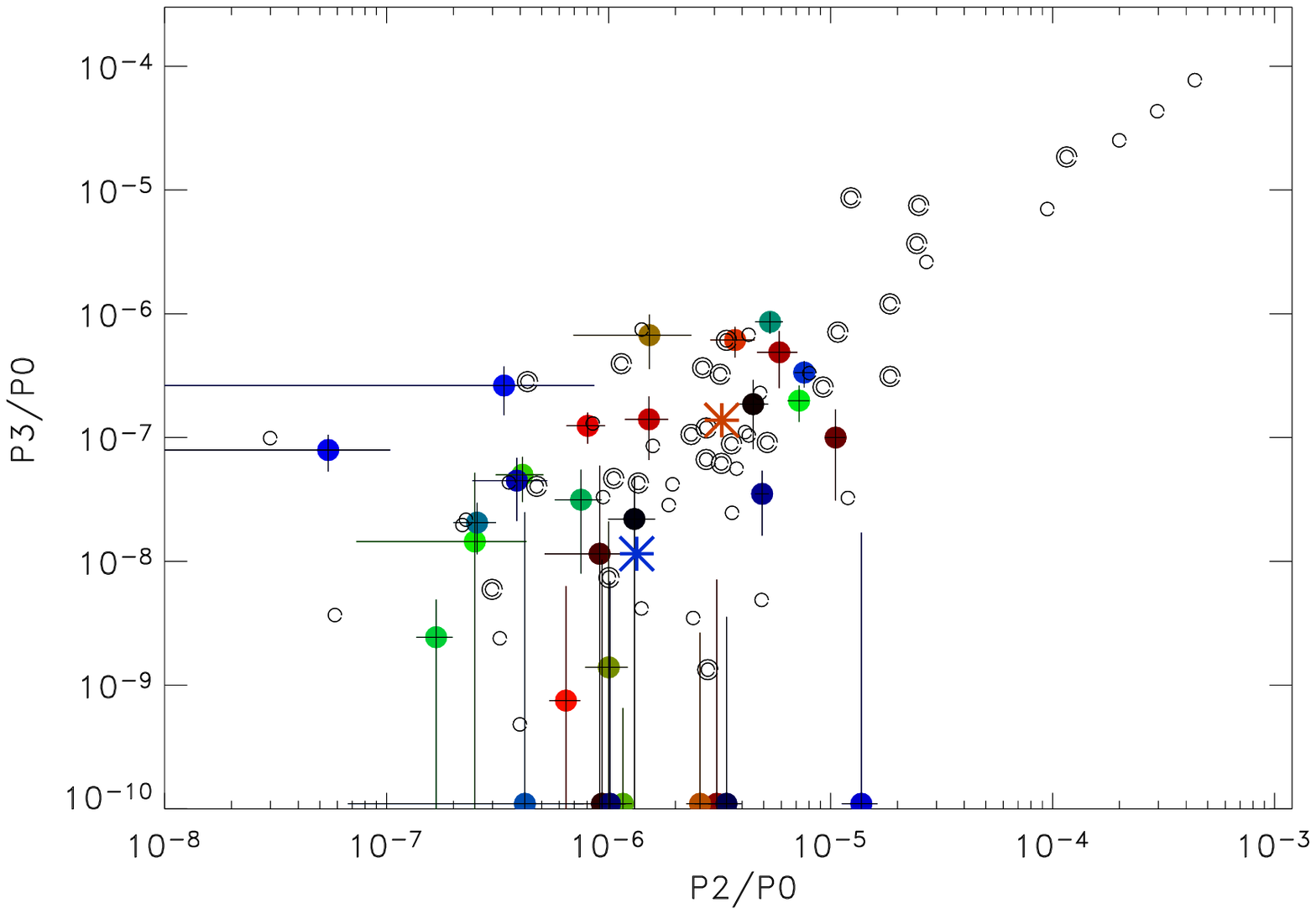}
      \caption{{\it Upper panel:} Comparison of power ratio parameter distributions of $P_2/P_0$ and $P_3/P_0$ 
      for the observed and simulated clusters. The lower, blue cross indicates the log mean of the 
      observed values and the upper crosses the values for the simulated clusters. Of these, the
      upper orange symbol is for all clusters and the slightly lower dark red mark for all 
      simulated clusters with temperatures above 2 keV. {\it Lower panel:} Same plot but now only 
      a subsample of 54 clusters from the simulations is shown that has a very similar ICM temperature 
      distribution as the observed clusters. Clusters with temperatures above 4 keV are marked with larger double 
      circles. In both plots we treat simulated galaxy clusters seen from different orthogonal viewing 
      angles as independent objects.}
  \label{Fig6}
  \end{center}
   \end{figure}

\subsection{Comparison of observations and simulations}

   \begin{figure}
   \begin{center}
   \includegraphics[width=6.5cm]{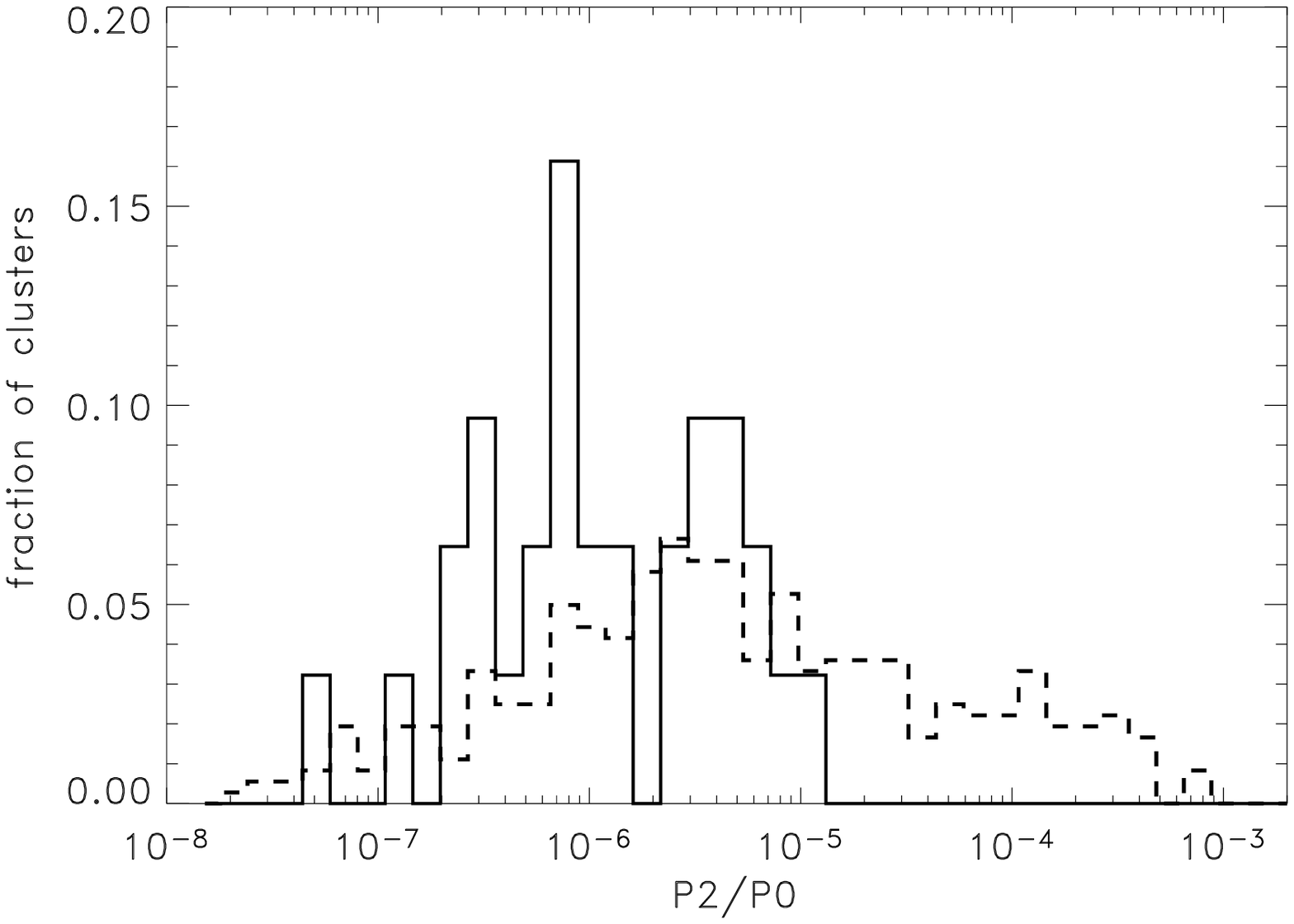}
   \includegraphics[width=6.5cm]{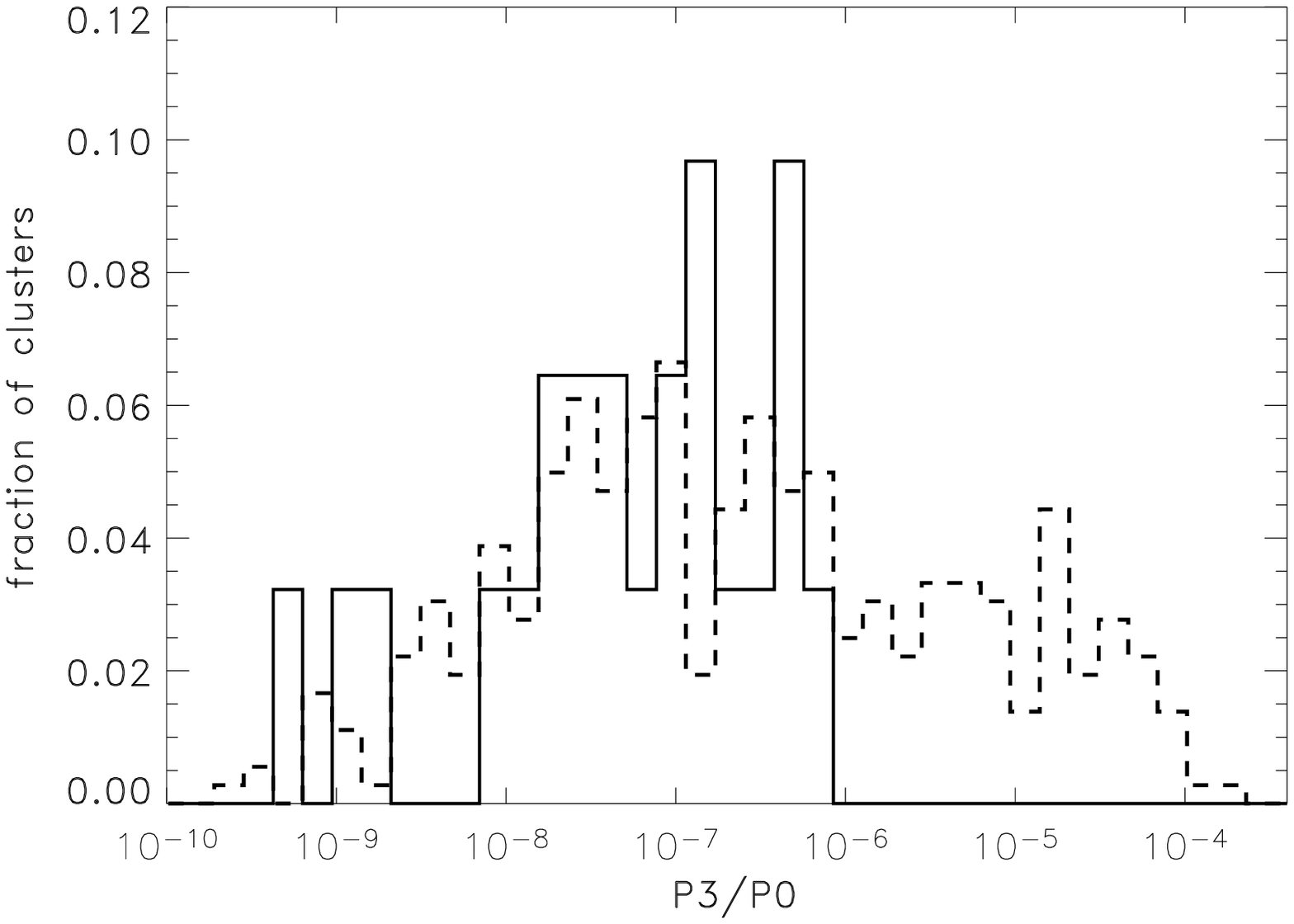}
   \includegraphics[width=6.5cm]{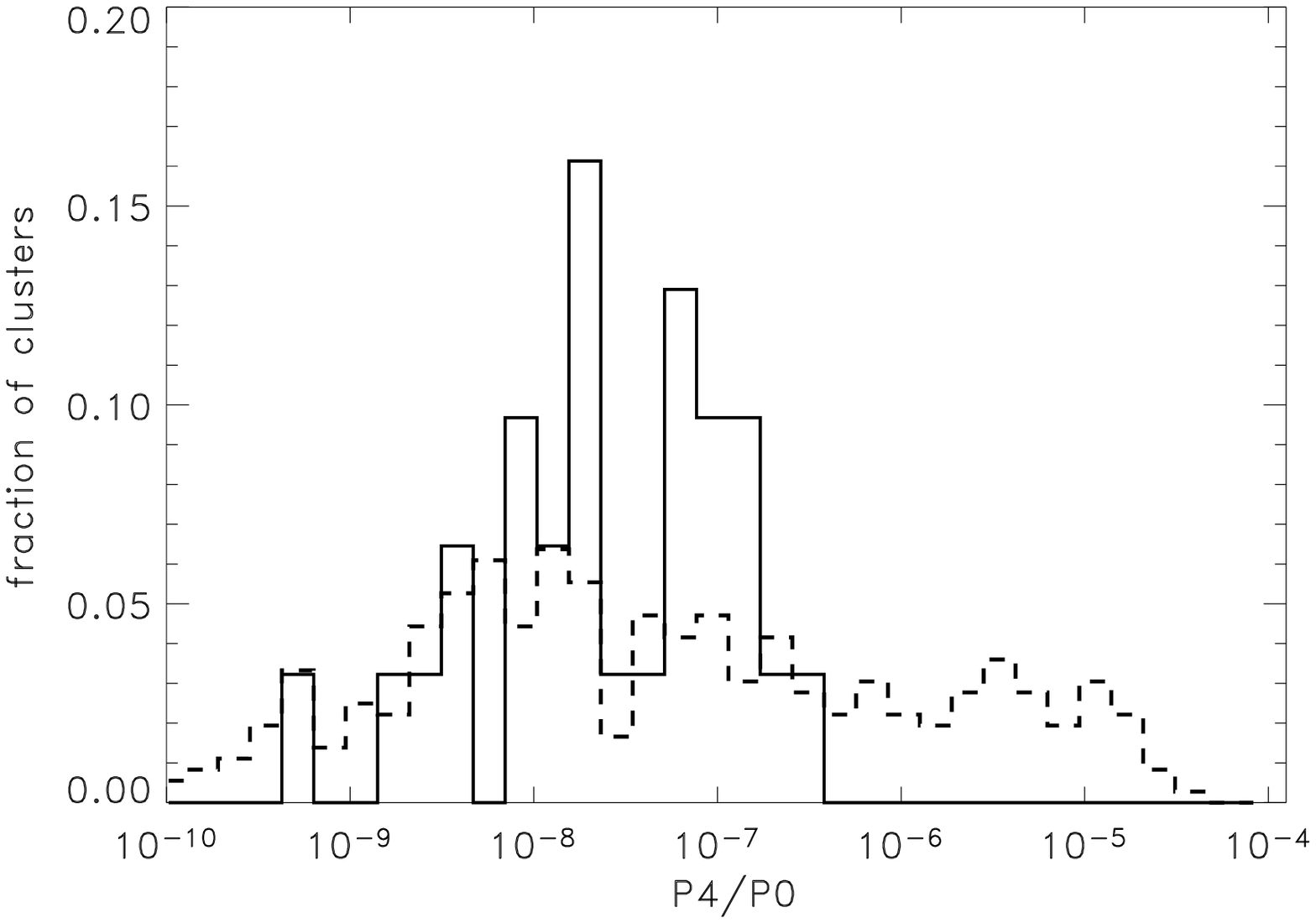}
      \caption{Comparison of the distribution functions of, from top to bottom,  
     $P_2/P_0$, $P_3/P_0$ and $P_4/P_0$, for the simulations (dashed lines) and 
     observations (solid line). All histograms are normalised by the total number 
     of clusters of each sample.}
  \label{Fig7}
  \end{center}
   \end{figure}
   \begin{figure}
   \begin{center}
   \includegraphics[width=7cm]{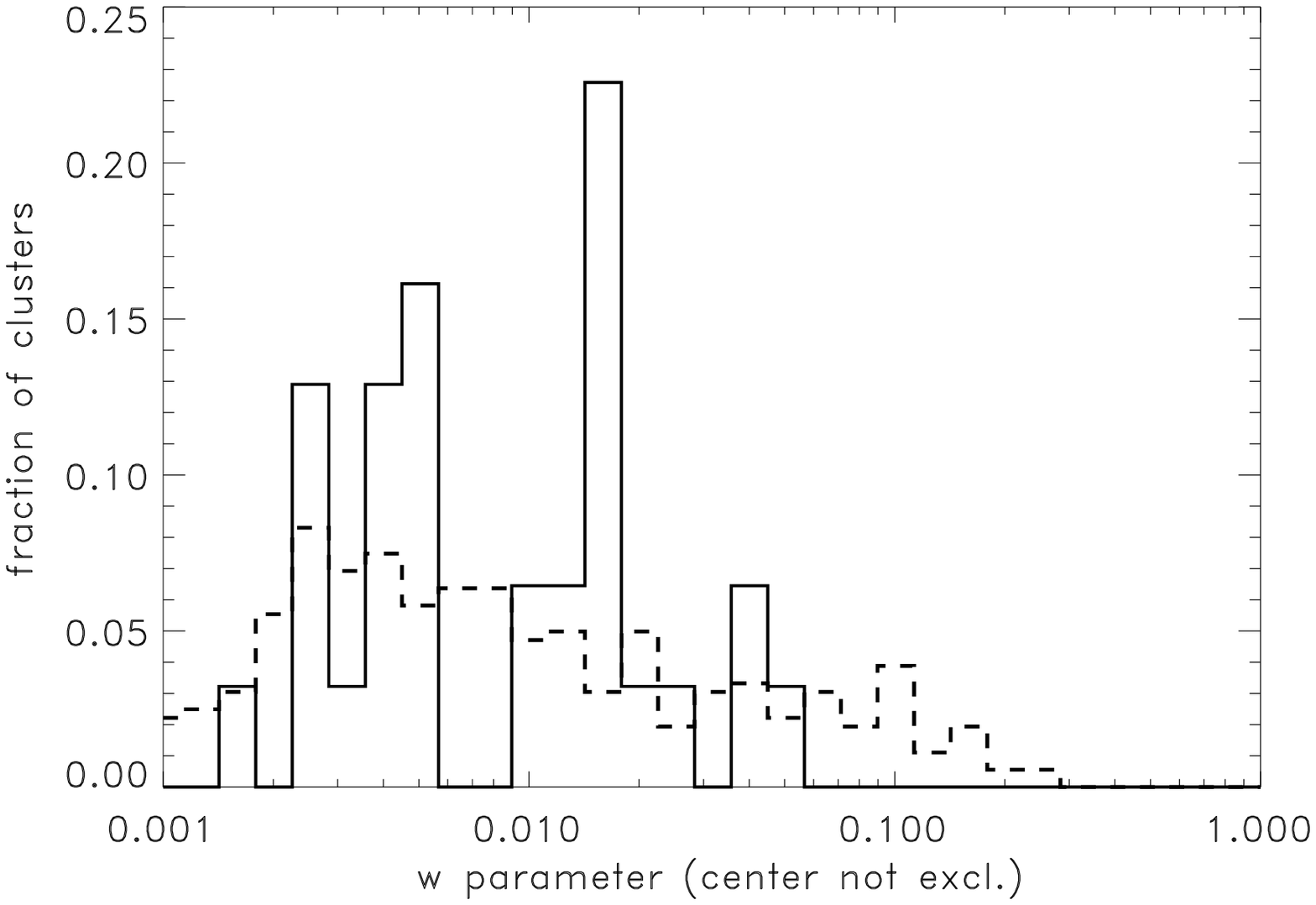}
   \includegraphics[width=7cm]{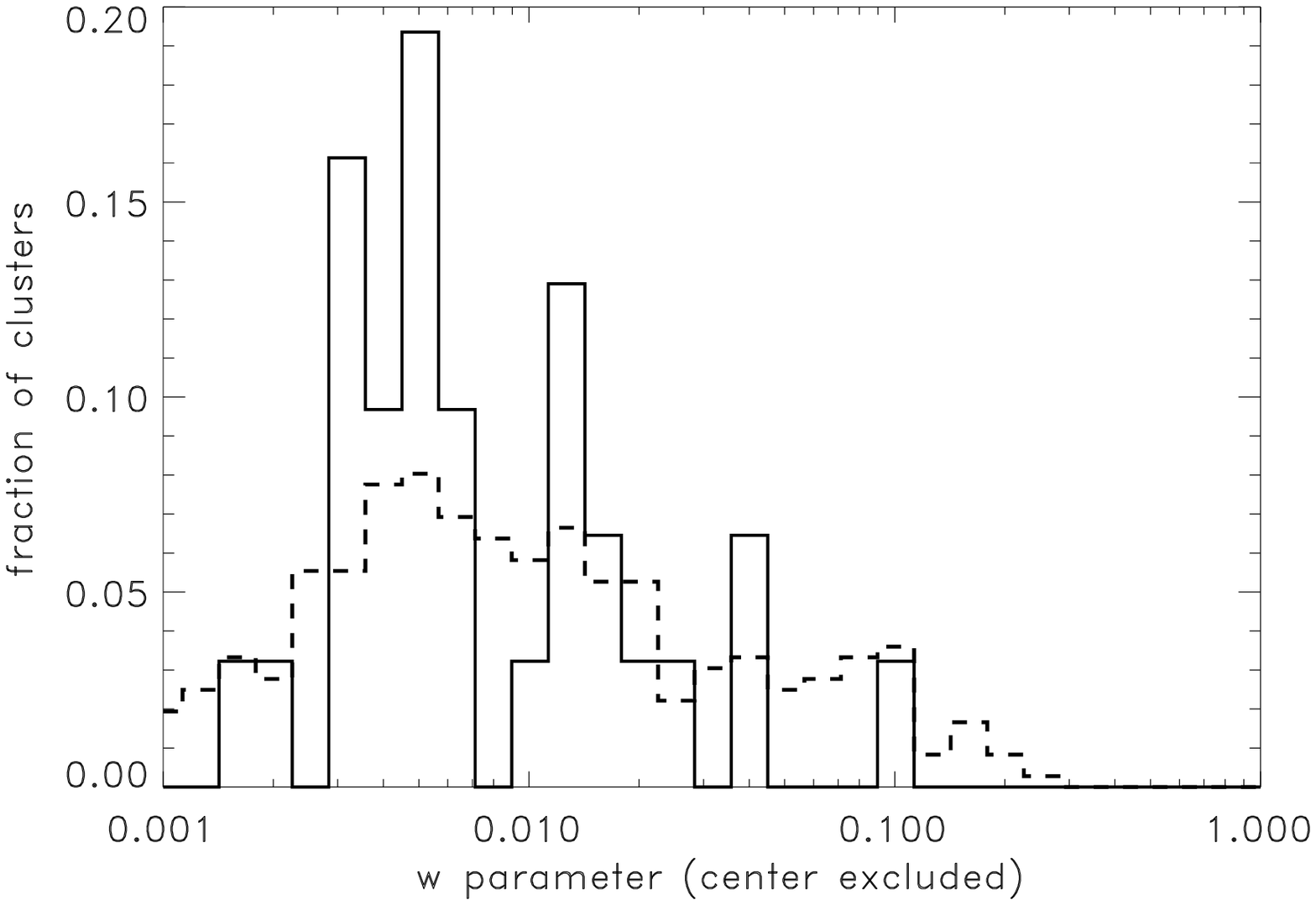}
      \caption{Comparison of the distribution functions for the $w$-parameter
obtained from the full aperture (top) and with central region excluded (bottom),  for the simulations (dashed lines) and observations (solid line). Both histograms are normalised by the total number of clusters in each sample.}
   \label{Fig8}
  \end{center}
   \end{figure}
 
As mentioned earlier, the simulation images do not contain photon noise, so that the substructure 
parameters we obtain have no statistical error and we also do not subtract a  photon noise bias. 
Figure~6 shows a comparison of the distribution of $P_2/P_0$ and $P_3/P_0$ for the simulations and 
observations. Since the simulation sample contains a large number of low temperature clusters outside 
the selection interval of \rexcess\ we have marked the clusters with temperatures above 2 keV with 
larger symbols. There is no apparent difference in the parameter distribution of the simulated 
clusters at $T_X \le 2$ keV and  $T_X \ge 2$ keV. While a larger fraction of the simulated 
clusters cover a similar parameter space to the observed objects in $P_2/P_0$ and $P_3/P_0$, 
there is a substantial fraction of simulated clusters with much higher substructure measures 
than \rexcess. To show this more quantitatively we have
determined the log-mean of the different distributions, as shown in the Figure~6. The log-mean 
of the observed clusters is at much lower values in both parameters. We also show the log-mean 
parameter value for the simulated galaxy clusters selecting only those systems with $T_X \ge 2$ keV, 
finding that the result does not differ significantly from that of the total sample. We have also 
checked that there is no significant difference using only clusters at $T_X \ge 4$ keV. The result 
that the substructure measures are largely independent of the cluster temperature when applied to 
the simulations supports the view that the discrepancy in mean values between the simulations and 
observations is not due to a mass or ICM temperature selection effect. We further corroborate these 
results with the analysis discussed below and shown in Fig. 15. Fig. 7 shows the histograms of all 
the power ratio values for the observed and simulated systems, underlining the fact that the power 
ratios of the simulated systems extend to much higher values than the observed objects. Fig. 8 shows 
the histograms of the $w$ parameter determined with and without excision of the centre. 
The discrepancy is more subtle in $w$ than for the power ratios, but is still significant.

   \begin{figure}
   \begin{center}
   \includegraphics[width=\columnwidth]{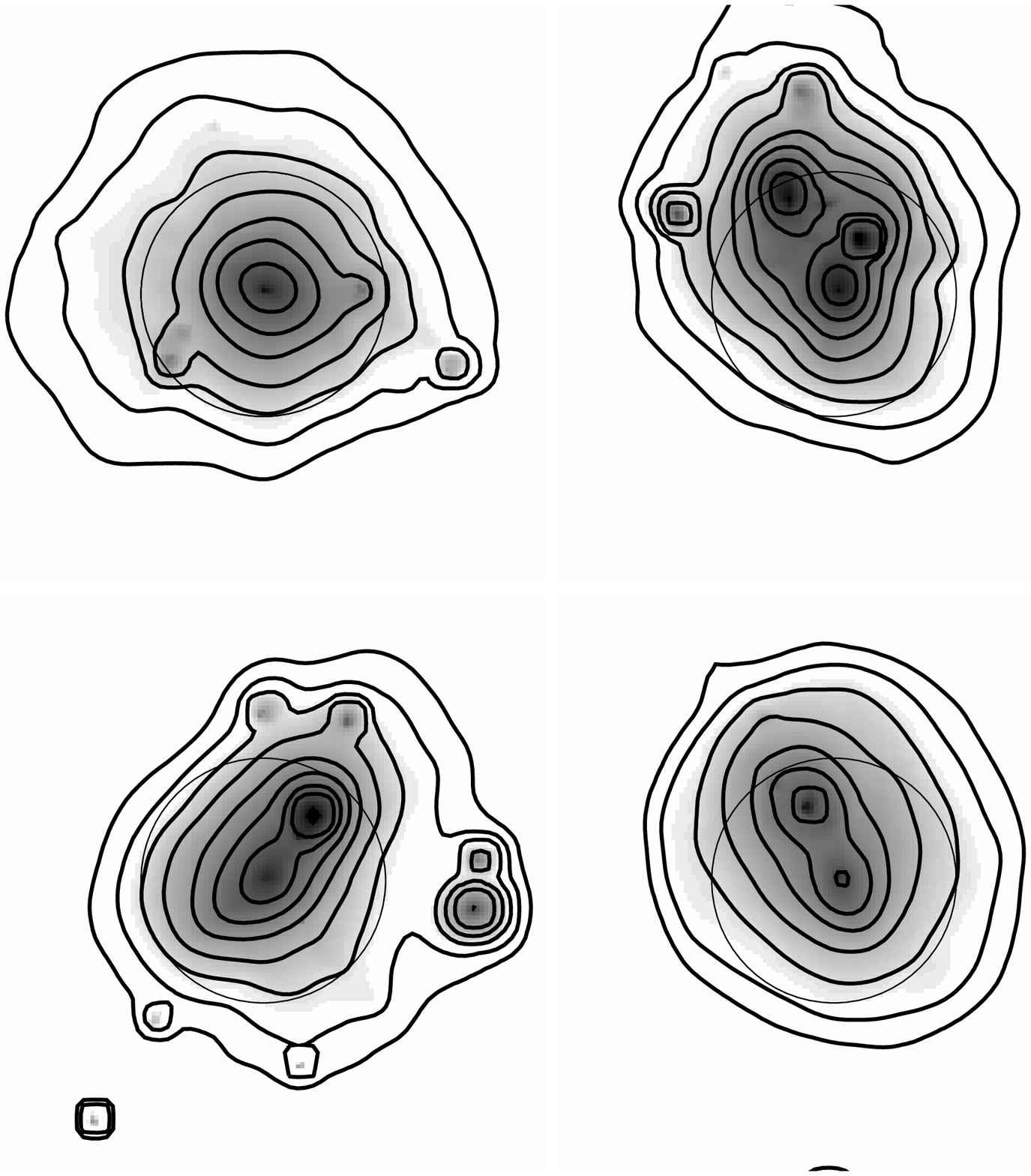}
      \caption{Examples of the four simulated clusters from Borgani et al. (2004) having the largest values of $P_3/P_0 > 10^{-4}$. The thin circle indicates a radius of $R_{500}$. All of these clusters show several clear maxima or clumps inside $R_{500}$, in contrast to the images of the \rexcess\ clusters. The images show only the diffuse emission from the ICM and the compact emission regions are not point sources but are in fact small cool cores.}
         \label{Fig9}
  \end{center}
   \end{figure}

Possible selection effects due to the different temperature and mass ranges covered by 
the simulated and observed cluster samples remains a major concern (e.g. 90 per cent of 
the simulated clusters have a temperature below 4.3 keV but only 55 per cent of the 
observed clusters have temperatures below this value). We thus performed another test 
to show that the excess of simulated clusters with strong indications for substructure 
is not due to selection effects. We resampled the simulated clusters in such a way that 
the distribution in temperature is roughly similar to the observed distribution. 
The resampling is not exactly perfect, since we have only 3 clusters with 3 viewing 
angles each in the temperature range from 4.3 to 6.5 keV and so we restore the balance 
by having more objects in the neighbouring bins. In total we compare 54 resampled clusters 
(treating the different viewing angles of the same cluster as independent values) 
to the 31 observed clusters with very similar temperature distributions in the lower 
panel of Fig. 6. We note that 22 per cent  (12/54) of the simulated clusters have power 
ratios in excess of the regime covered by the observed clusters. We also use different 
symbols for simulated clusters below and above 4 keV and note that, even given the small 
number, hotter and cooler clusters have similar power ratio distributions. Therefore we 
are confident that we can rule out that the discrepancy in the power ratio distributions 
between observed and simulated clusters is due to a selection effect.

A very similar result for the distribution of the power ratio parameters
is found when comparing with the simulations by Valdarnini (2006). The $P_3/P_0$ of 
his simulated clusters span the range of $10^{-8}$ to $10^{-4}$,
and thus these simulations also populate the parameter range from $10^{-6}$ to 
$10^{-4}$, in which there are no observed clusters.

In search of a physical reason for the difference in morphological statistics between 
simulated and observed clusters, we inspected the images of the simulated objects with 
large substructure parameters. Fig.~9 shows four examples of simulated clusters from the 
extreme upper right corner of Fig.~6. These images contain only the diffuse X-ray emission 
of the ICM, and what may appear to be point sources are very compact cool regions that have 
been accreted by the clusters. We have marked in the Figure the aperture radius $R_{500}$ 
inside which the substructure analysis is undertaken. All of the simulated clusters show 
noticeable substructure features inside $R_{500}$ that serve to boost the power ratios, 
in particular if they are located close to the aperture radius. In the real cluster images 
we do not find equivalent compact emission regions. It thus appears that one significant 
difference between observations and simulations is the fact that at least a fraction 
of the simulated clusters contain more compact cool cores than their observed counterparts.

Pratt et al. (2007), when comparing the temperature profiles of the \rexcess\ with those from 
this same sample of simulated clusters, showed that almost all simulated objects had a central 
temperature decrease signifying the presence of a pronounced central cool core, whereas 
less than half of the observations showed this feature. From this finding one could have 
expected that simulated clusters are more regular on average, since cool core clusters 
have statistically less substructure than non-cool core clusters. The explanation is 
more complex. The simulated clusters not only have more pronounced cool cores in their 
central regions, but they also contain previously accreted subclusters that themselves 
have strong cool cores. These survive in the final cluster and produce multiple maxima, 
as seen in Fig.~9. We can illustrate the overabundance of cool regions in the simulations 
with another statistic from our analysis. If we characterize the strength of a cool core 
by $L_{\rm rat}$, the ratio of the cluster flux from the total cluster image interior 
to $R_{500}$ to that with the core region ($r < 0.1\, R_{500}$) excised, we find, 
as shown in the histogram in Fig.~10, that the simulations cover a somewhat broader 
range of such flux ratios, extending  up to higher values than the observations. 

   \begin{figure}
   \begin{center}
   \includegraphics[width=7cm]{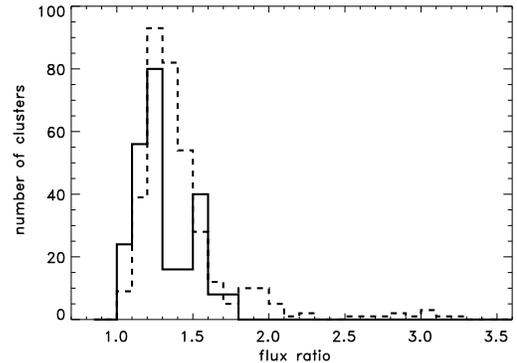}
      \caption{Comparison of the ratio of the total cluster flux to the flux
      with the core region ($r < 0.1\, R_{500}$) excised for 31 clusters from \rexcess\ (solid line) and the simulated clusters (dashed line).}
         \label{Fig10}
  \end{center}
   \end{figure}

\section{Correlation of morphological and global cluster parameters}

   \begin{figure}
   \begin{center}
   \includegraphics[width=\columnwidth]{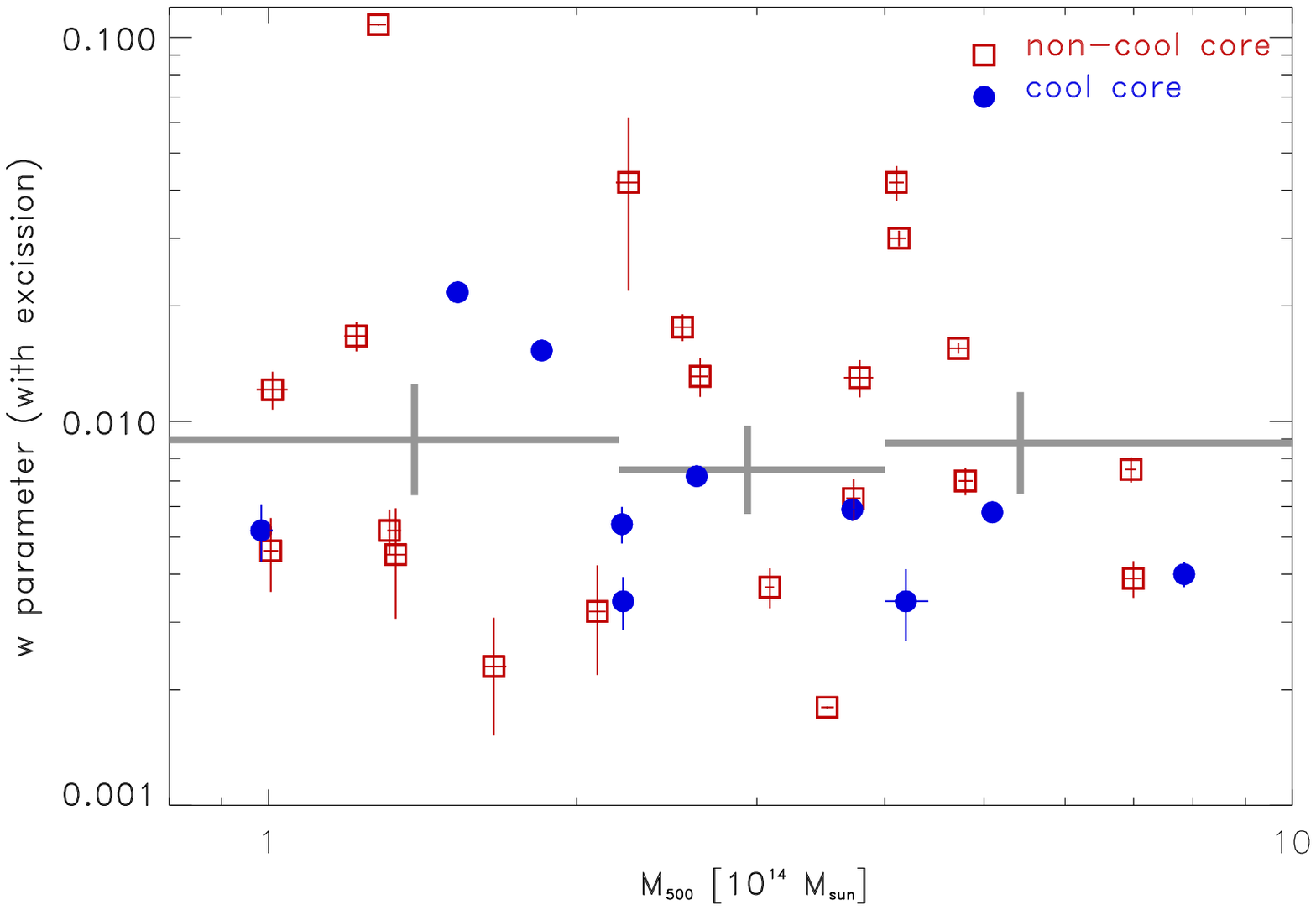}
   \includegraphics[width=\columnwidth]{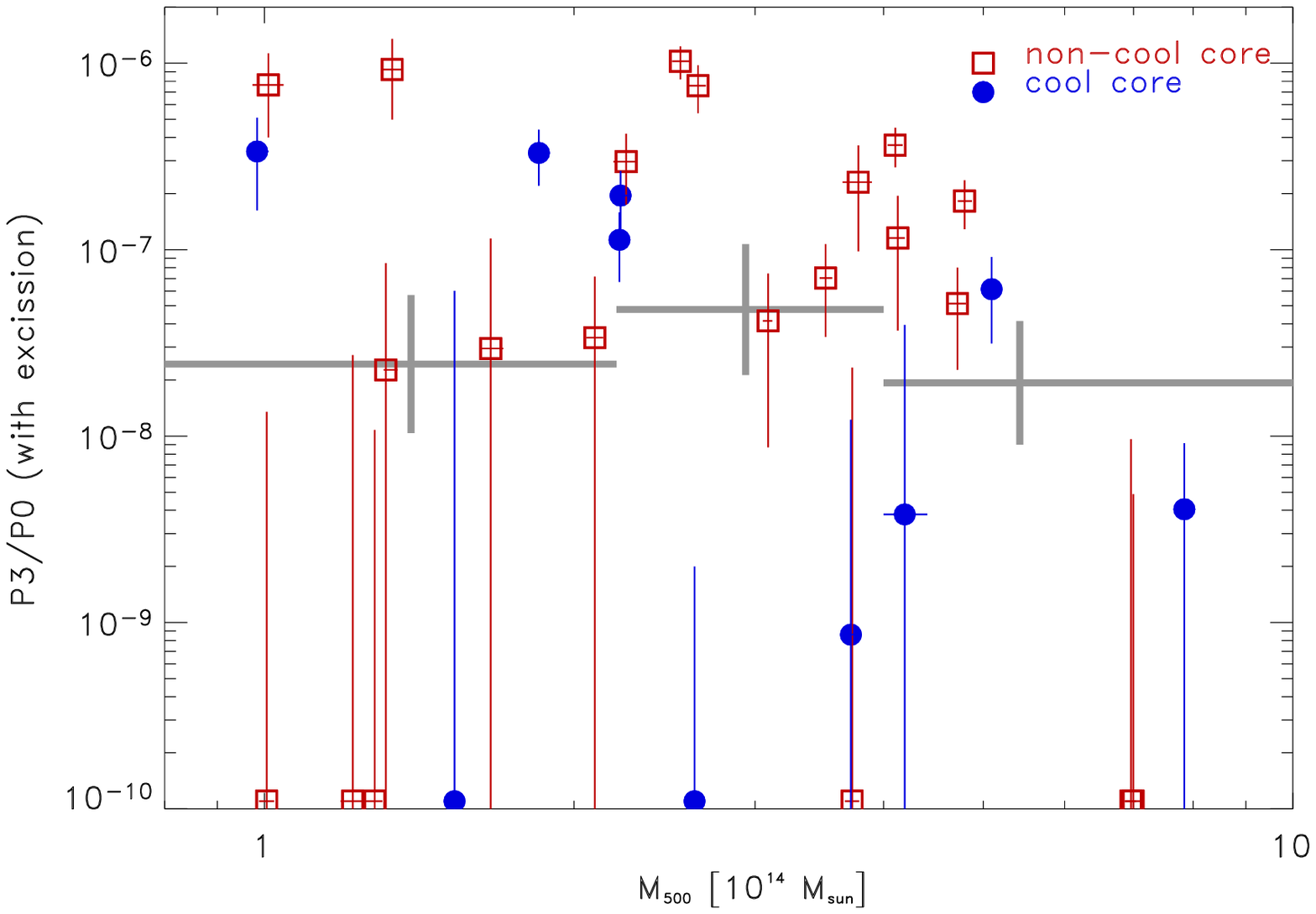}
      \caption{Correlation of the substructure parameter $w$ (top) and $P_3/P_0$ (bottom) 
      with cluster mass for the 31 cluster from \rexcess\ . The grey bars show the log-mean 
      values of $w$  and $P_3/P_0$ determined for three mass bins.}
         \label{Fig11}
  \end{center}
   \end{figure}

   \begin{figure}
   \begin{center}
   \includegraphics[width=\columnwidth]{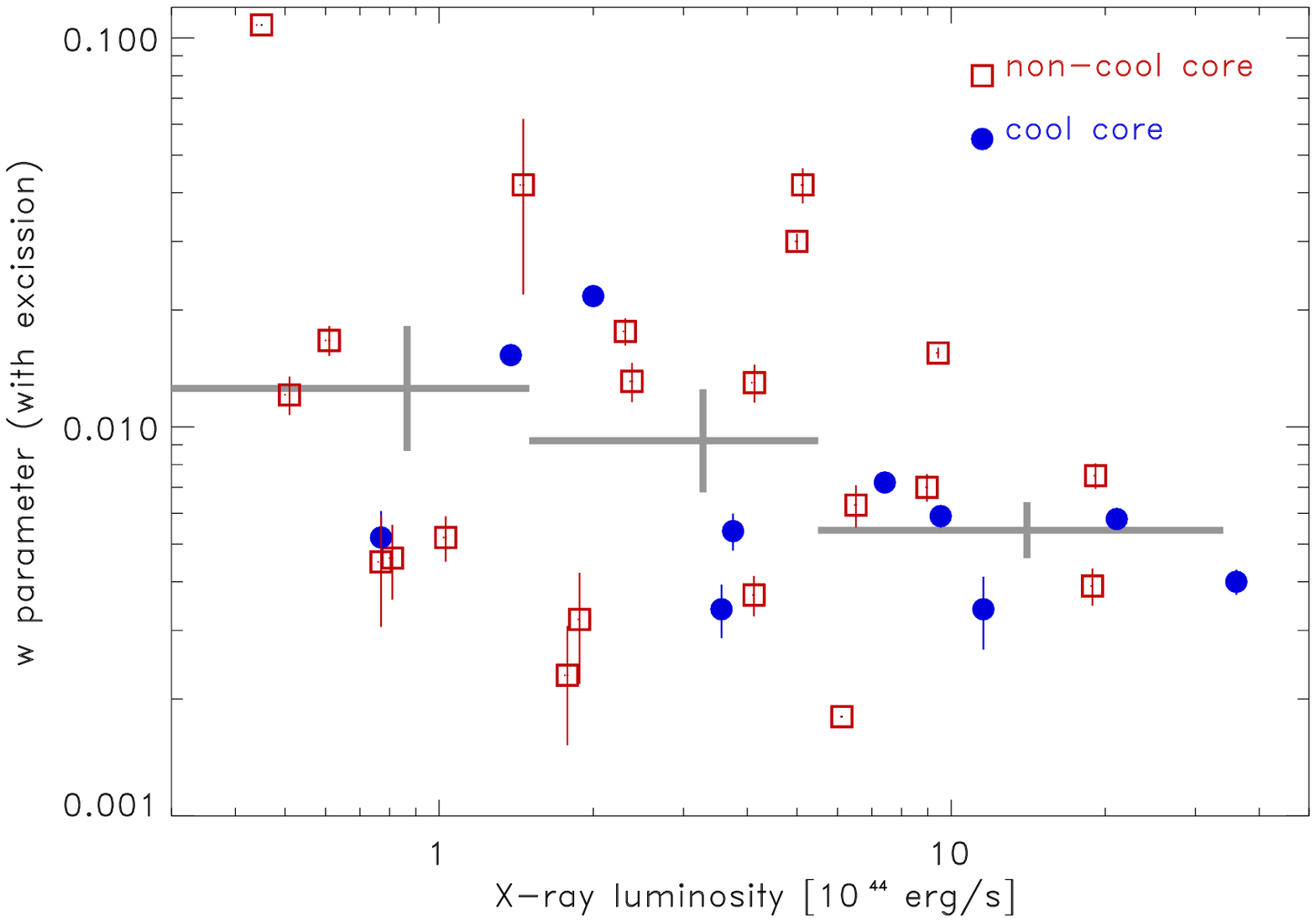}
   \includegraphics[width=\columnwidth]{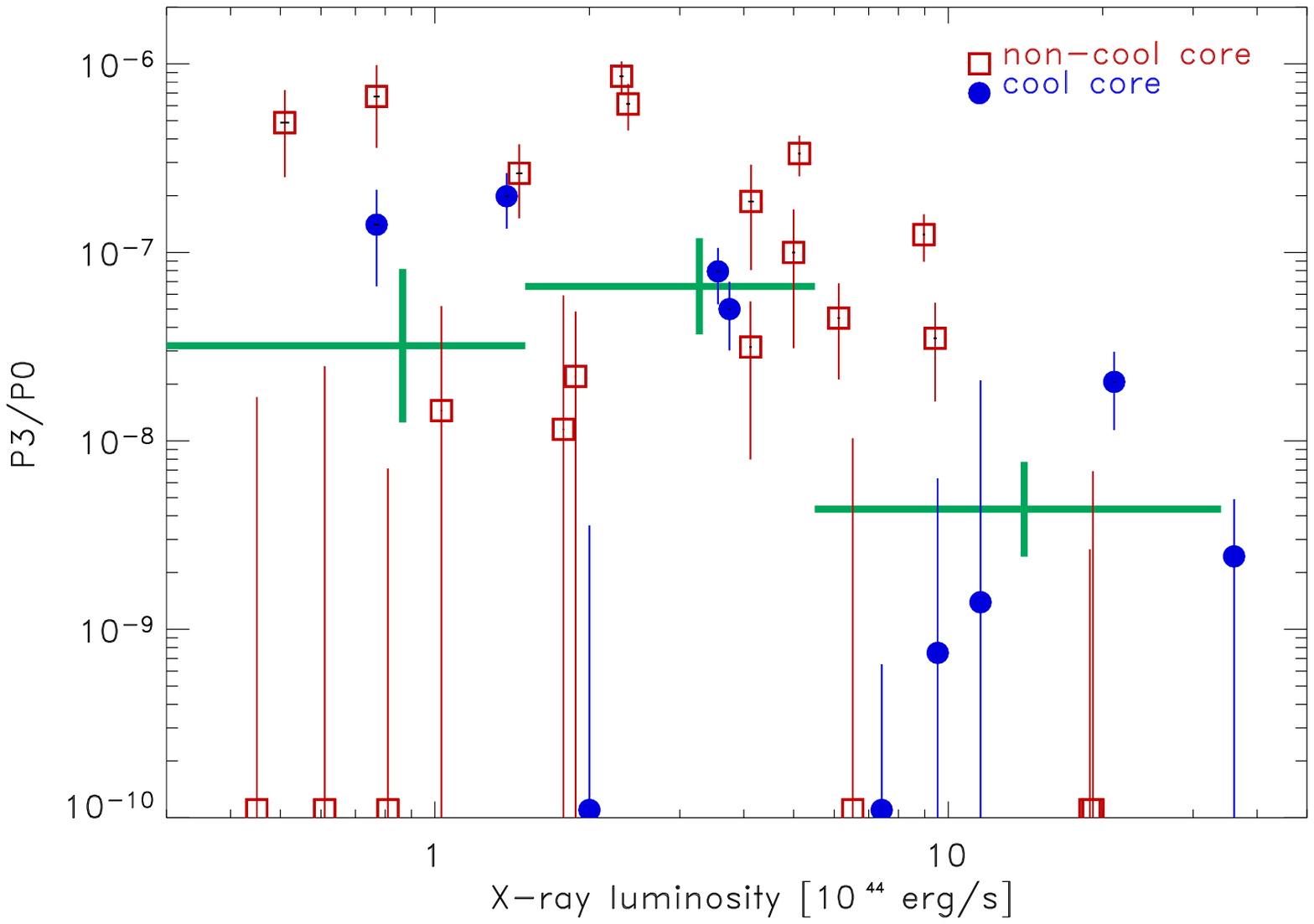}
      \caption{Correlation of $w$ (top) and $P_3$ (bottom), both obtained with the core regions excised, with bolometric X-ray luminosity estimated interior to  $R_{500}$ aperture (Pratt et al. 2009a). The grey bars show the log-mean values of $w$ determined for three luminosity bins.}
         \label{Fig12}
  \end{center}
   \end{figure}

   \begin{figure}
   \begin{center}
   \includegraphics[width=\columnwidth]{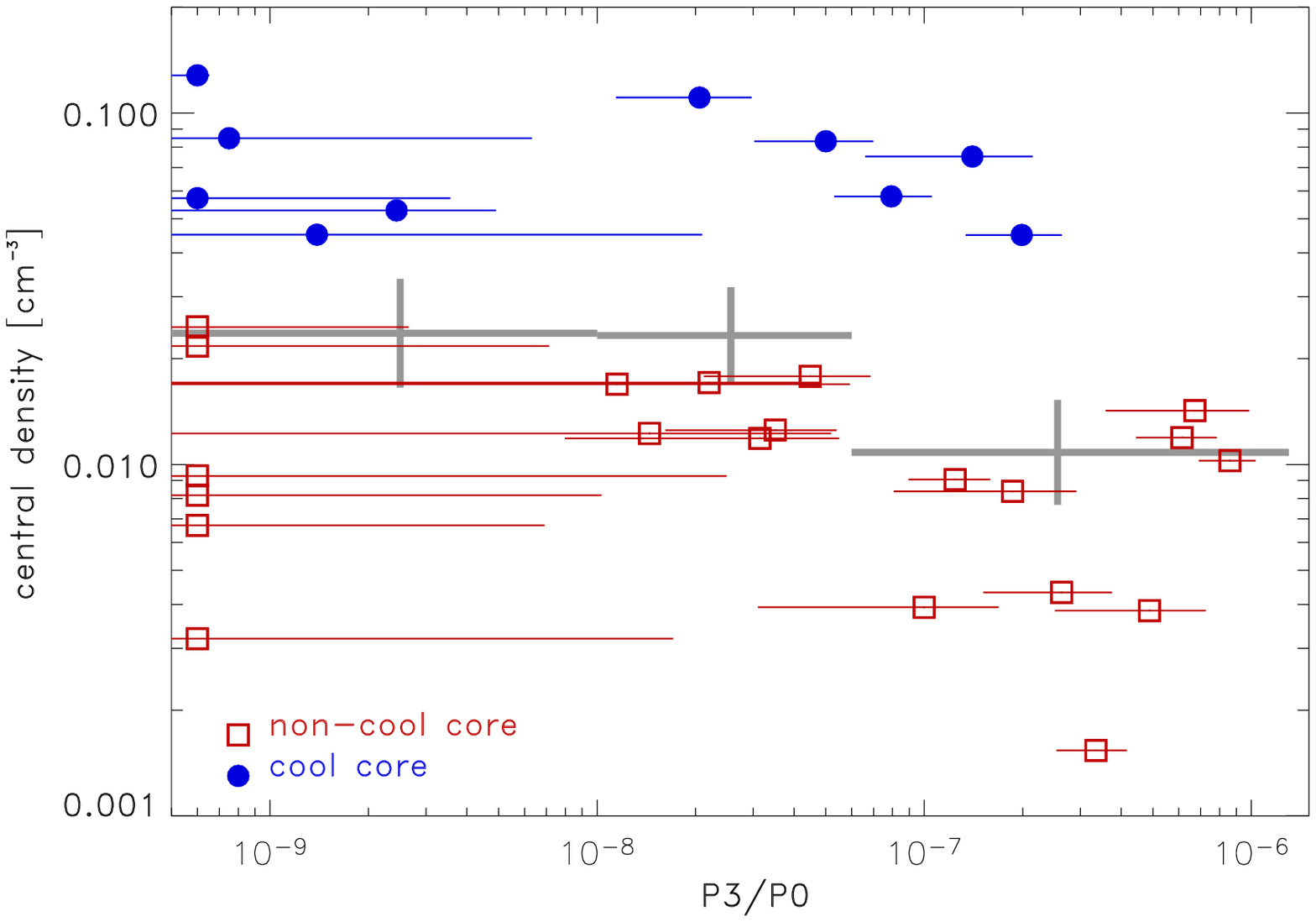}
   \includegraphics[width=\columnwidth]{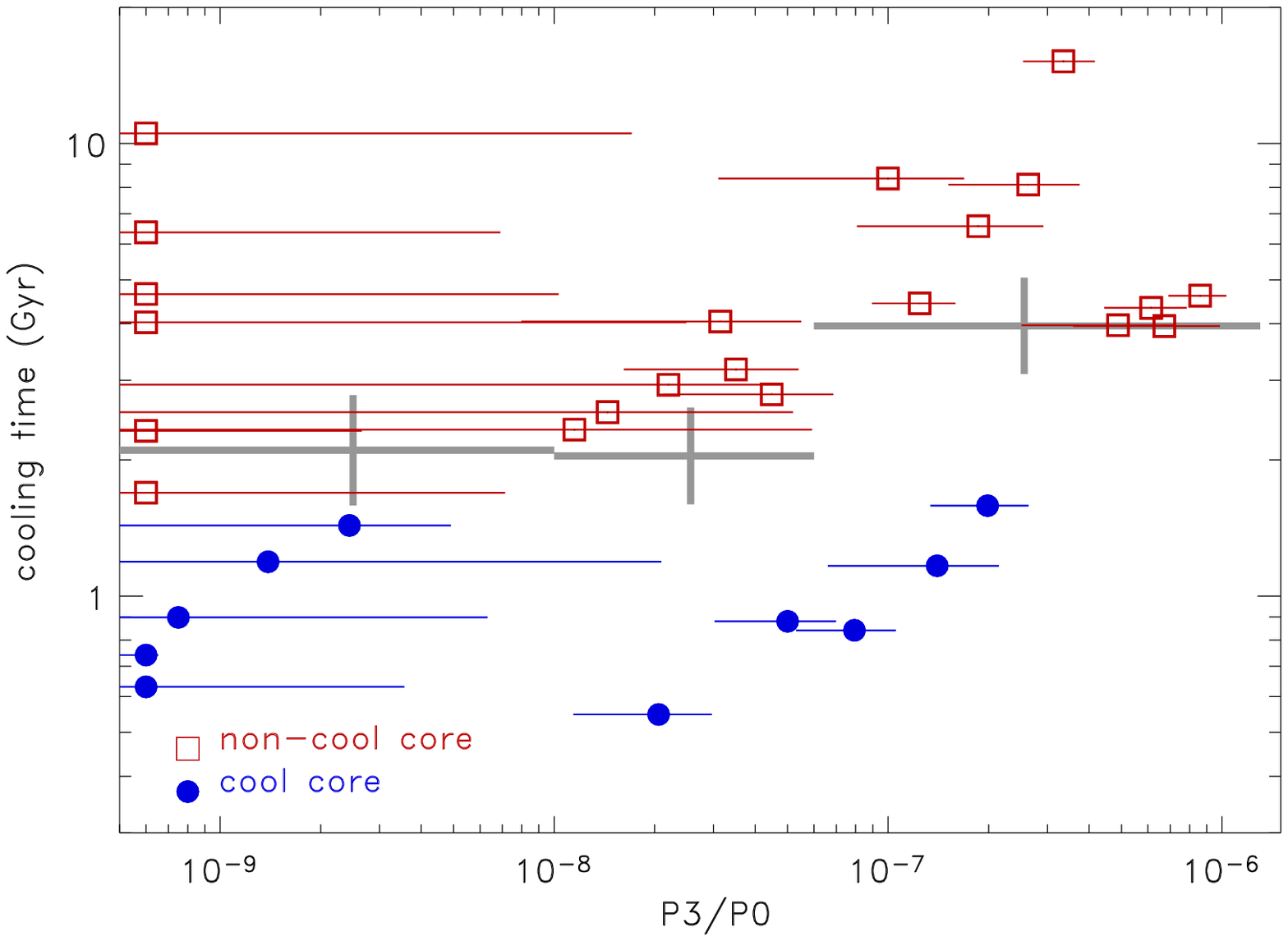}
      \caption{Correlation of the central ICM density (top) and cooling time (bottom) with the $P_3/P_0$. Similar correlations between the centre shift parameter $w$ and the central density and cooling time were shown in Figure~12 of Croston et al. (2008).}
         \label{Fig13}
  \end{center}
   \end{figure}
   \begin{figure}
   \begin{center}
   \includegraphics[width=7.5cm]{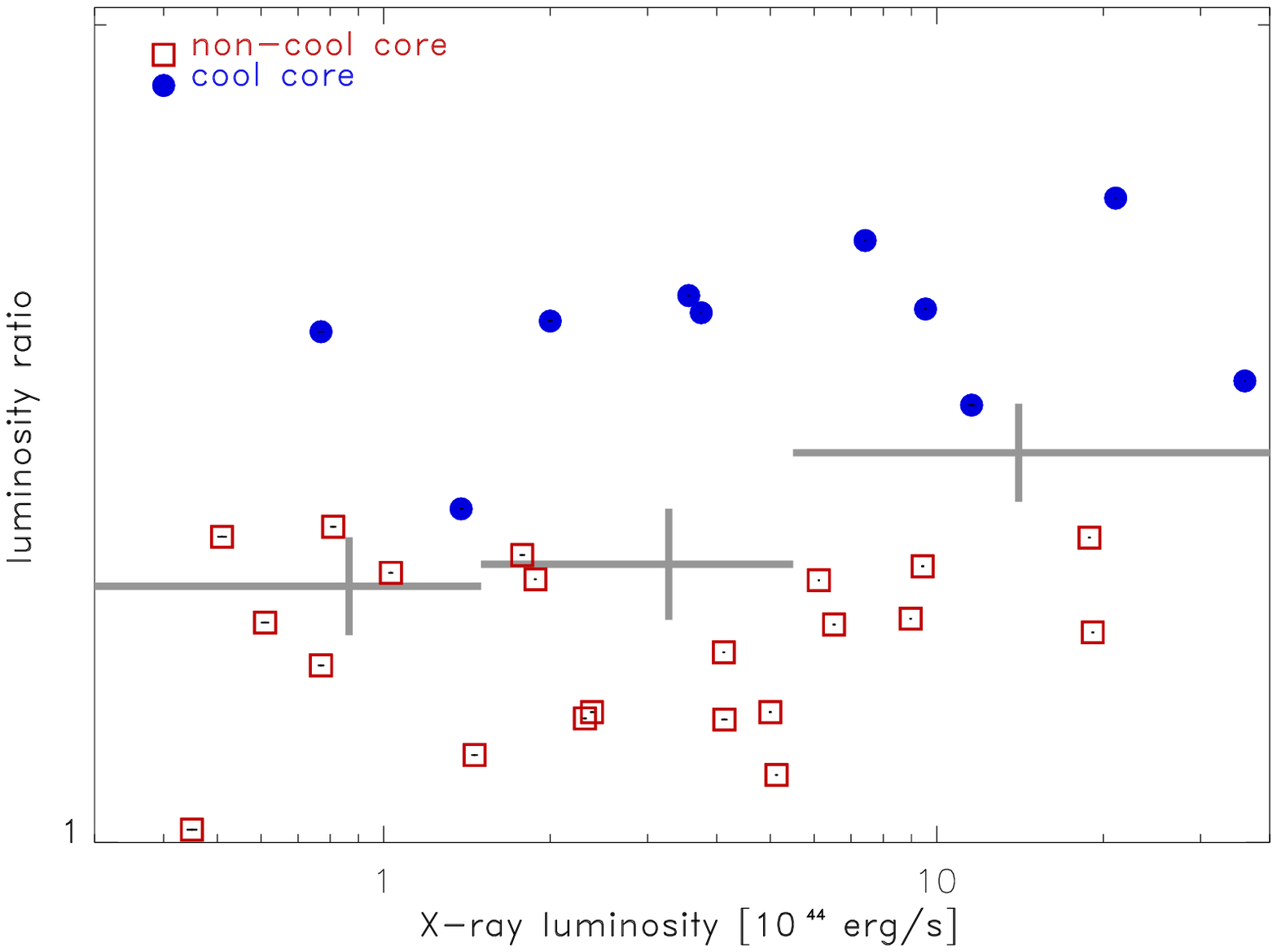}
   \includegraphics[width=7.5cm]{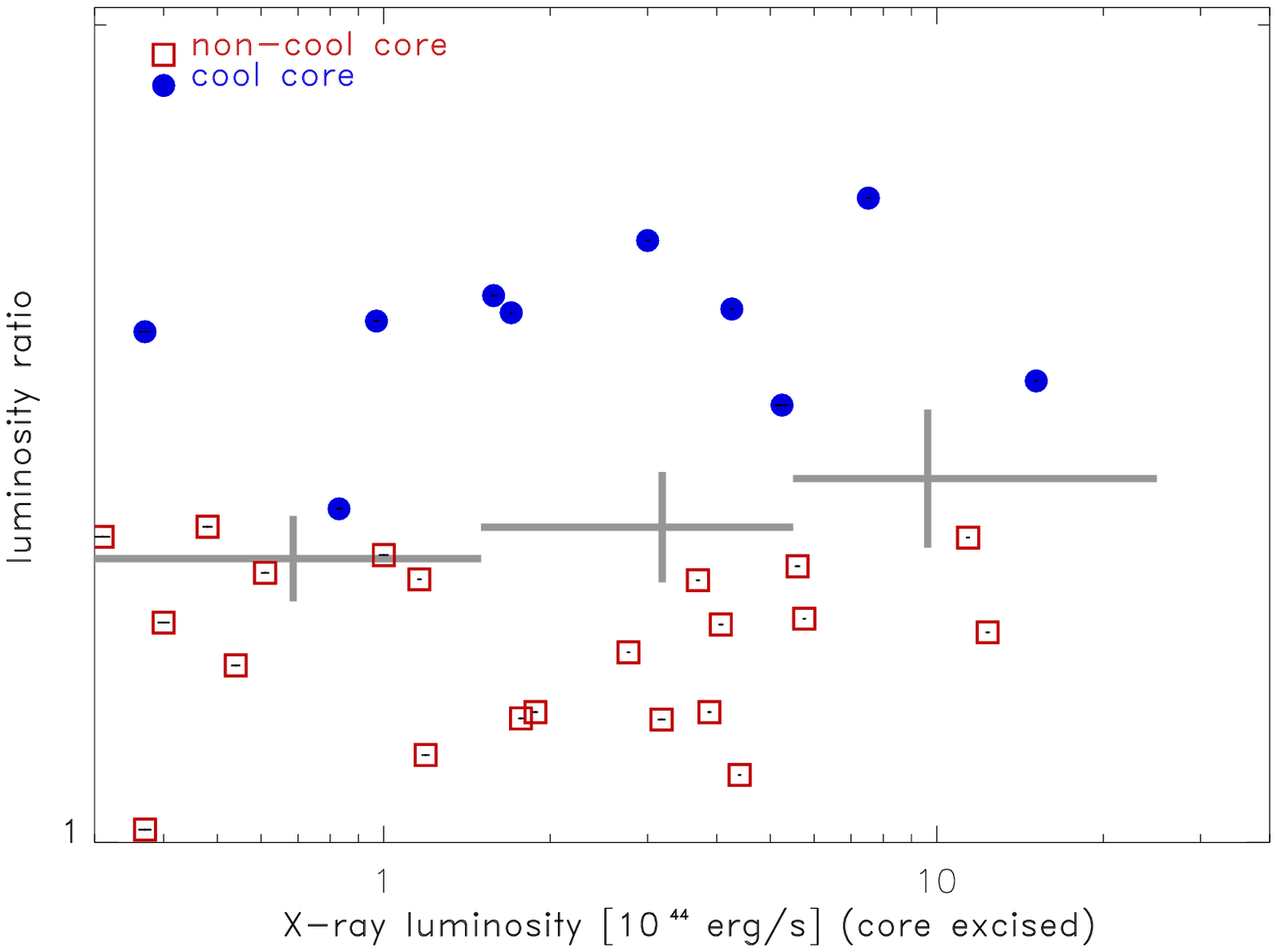}
   \includegraphics[width=7.5cm]{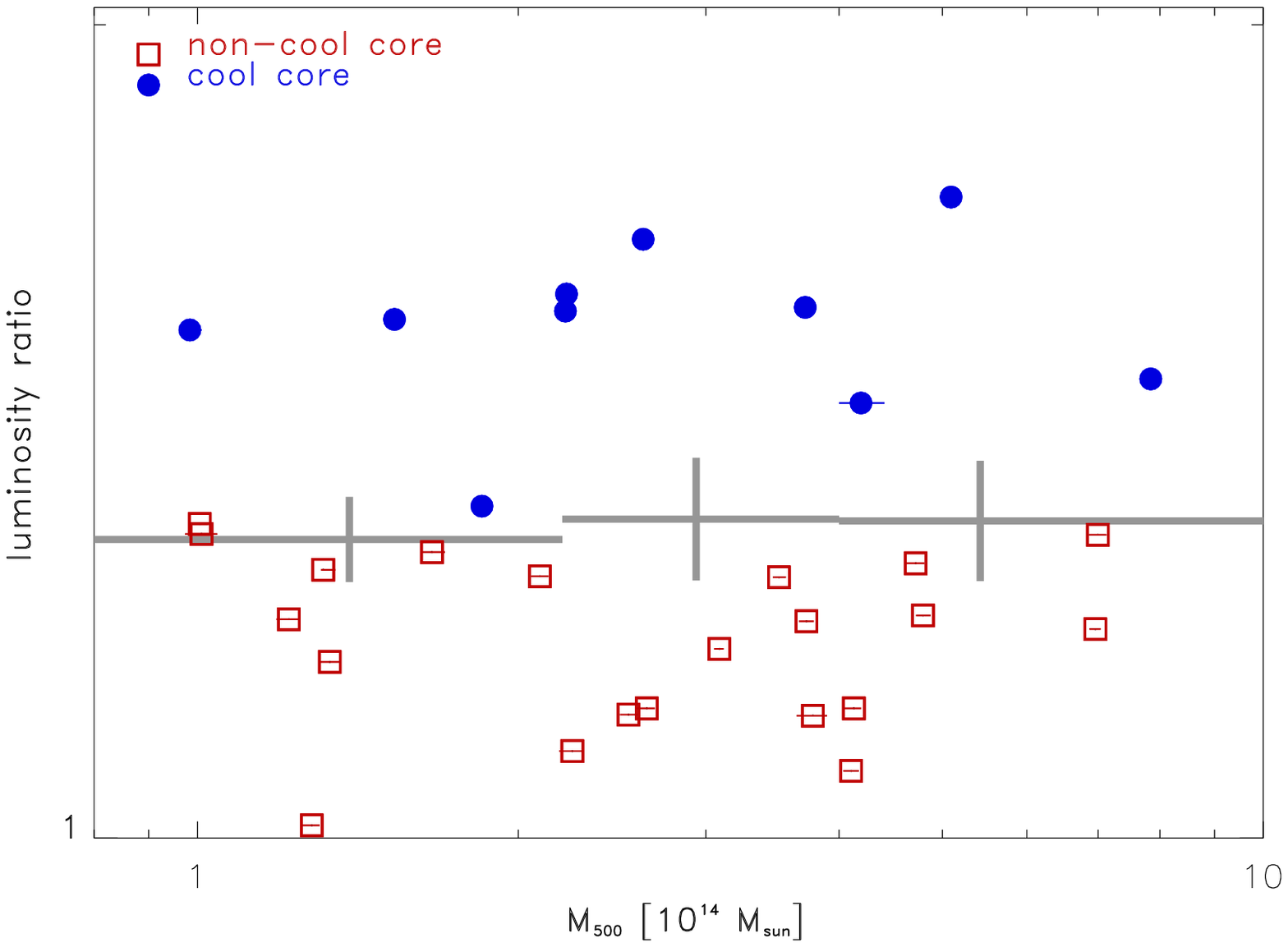}
      \caption{Correlation of the bolometric X-ray luminosity interior to $R_{500}$ (top), core excised bolometric X-ray luminosity (middle), and mass (bottom) with the X-ray luminosity ratio $L_{\rm rat}$. The latter is defined as the ratio of the total flux in the [0.5-2] keV band measured interior to $R_{500}$ to the flux in the same aperture with the core region ($r < 0.1\, R_{500}$) excised. This luminosity ratio closely correlates with other cool core properties such as  central density and cooling time, and is a sensitive indicator of cool cores. The grey bars show the log-mean values of $L_{\rm rat}$ determined for three bins.}
         \label{Fig14}
  \end{center}
   \end{figure}

In this Section, we investigate how the substructure measures vary with global cluster 
properties. We start by investigating how the substructure parameters vary with mass, 
since this is the most fundamental scaling parameter of a cluster. We use the \rexcess\ 
mass estimates given in Pratt et al. (2009b), which were obtained from iteration about 
the $M_{500}-Y_X$ relation. Figure~11 shows $w$ and $P_3/P_0$, obtained with central 
region excised, as a function of mass. Also overplotted are logarithmically averaged 
values in three mass bins. There is no obvious variation in the occurrence and strength 
of substructure with cluster mass, a result that is quantitatively confirmed by the statistical 
tests listed in Table~3. For $w$, a Kendall's $\tau$ test gives a probability of 0.81 and 
a Spearman's rank test a probability of 0.81 for no correlation. For $P_3/P_0$, the 
corresponding probabilities are 0.69 and 0.73, clearly pointing towards no mass correlation.

Next we use bolometric X-ray luminosity, $L_X$, as the scaling parameter, since this is the most frequently used observable. Figure~12 shows $w$ and $P_3/P_0$ as a function of the \rexcess\ $L_X$ values published by Pratt et al. (2009a), with averages in three $L_X$ bins overplotted. The Kendall's $\tau$ and Spearman's $\rho$ probabilities are 0.63 (0.14) and 0.63 (0.22) for $w$ ($P_3/P_0$), respectively, suggesting that at least for the correlation of the power ratios with X-ray luminosity the observed weak correlation is statistically confirmed.

We further investigate how substructure and the cool core properties are connected. 
This was already partly explored in Croston et al. (2008, their Fig. 12) using 
correlations of $w$ and $P3/P0$ with central gas density (at 0.007 $R_{500}$) 
and central cooling time (at 0.03 $R_{500}$). Croston et al. found that these 
data allowed to reject the hypothesis of no correlation with probabilities of 
$\sim 85 - 92\%$. We show a similar analysis for the power ratios in Fig.~13. 
A Kendall's $\tau$ test gives a probability of no correlation of $12\%$ ($10\%$) 
and a Spearman rank test a probability of $6\%$ ($6\%$) for central density 
(cooling time), respectively. We have also studied the variation of $w$ with 
the central density and cooling time -- results are given in Table~3. The correlation 
with $w$ is even tighter than for $P_3/P_0$. In a study presented in section A.3 
in the Appendix, we show that also the strength of the correlation between 
$P_3/P_0$ and cool core indicators increases if we decrease the aperture radius, 
thus giving less weight to the very outer regions of the object.

There is thus evidence for a reasonably good correspondence between global morphological parameters and core properties. This quantifies for the first time in a representative sample the widely expected result that cool core systems correspond to clusters that have not been disturbed by mergers in the recent past. But the fact that the correlation is far from being perfect implies that the presence of a cool core can not generally be taken as an indication that a cluster is relaxed. On the other hand, it is worth noting that the correlation seen in Fig.~13 is all the more remarkable given that it involves $P_3/P_0$, which provides a measure of substructure on a very global scale with little influence from the central regions, as discussed above. This is the reason why we preferred to show this relation rather than the tighter correlation with the $w$ parameter. The correlation we find therefore demonstrates that there is a causal, statistical connection between the properties of the very central region and the global morphology.  



The above results provide the key to understanding the different correlations
between mass, X-ray luminosity and cluster morphology. We know that for a given mass, 
clusters with cool cores have in general higher X-ray luminosities (e.g. Fabian et al. 1994, 
Chen et al. 2007, Pratt et al. 2009a). Thus in going from the mass distribution to 
the $L_X$ distribution, cluster cool cores preferentially move to higher luminosities compared 
to non-cool core clusters. Since these cool core clusters are on average more regular, 
the more regular clusters will accumulate at the higher luminosity side -- exactly as observed. 
Thus, if we accept galaxy cluster mass as the primary scaling parameter, the correlation 
of $L_X$ with the substructure parameters can be seen as a selection effect.

To close the loop of arguments, we can also test our expectation that 
cool cores are preferentially found in the higher luminosity bins. 
Figure~14 shows the luminosity ratio $L_{\rm rat}$, defined as the ratio 
of the total flux in the [0.5-2] keV band measured interior to $R_{500}$ 
to the flux in the same aperture with the core region ($r < 0.1\, R_{500}$) excised. 
This parameter was defined to separate out cooling cores and is, as we have tested, 
very tightly correlated to central density and cooling time. There is a noticeable 
correlation and it is mostly the highest luminosity bin that features
a higher $L_{\rm rat}$ on average. The statistical tests listed in Table~3
show a probability of non correlation of 0.16 and 0.18 on a Kendall's $\tau$ 
test and a Spearman's rank test, respectively, supporting a significant
correlation. If the same test is done for the correlation of $L_{\rm rat}$ with
the cluster mass, as shown in the lower panel of Fig.~14, we find that it is 
rejected with probabilities of 0.97 and 0.84 on a Kendall's $\tau$ test and 
Spearman's rank test, respectively. In the middel panel of Fig. 14 we also show
the correlation of $L_{\rm rat}$ with the core excised X-ray luminosity
of the clusters. The no-correlation probabilities of 0.97 and 0.84 provided by 
Kendall's $\tau$ test and Spearman's rank tests indicate no significant correlation.
Thus, core excision in the luminosity integration removes the influence of
cool cores quite effectively.

\section{Discussion}

   \begin{figure}
   \begin{center}
   \includegraphics[width=\columnwidth]{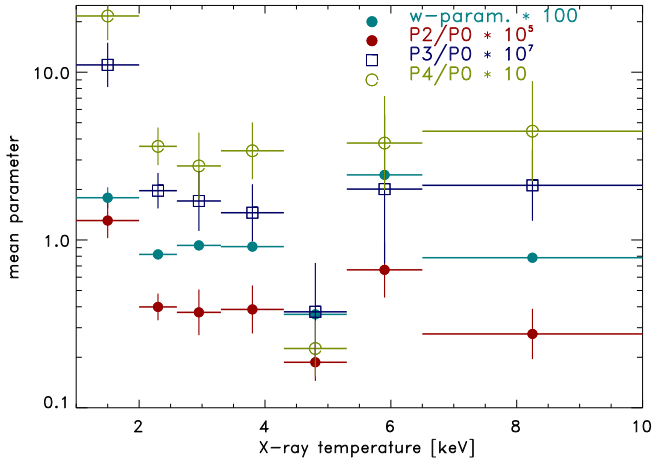}
      \caption{Substructure measures, power ratios and centershifts, for the simulated
      clusters as a function of ICM temperature. The substructure measures have been averaged
      for seven temperature bins of the simulated cluster sample. For each cluster all three
      projections have been used as independent results. Note that bin 5 contains only one and bin
      6 only two clusters.}
         \label{Fig15}
  \end{center}
   \end{figure}

\subsection{Methodology}

The results presented in this paper provide insight into the reliability and sensitivity of two methods to characterise substructure: power ratios and centre shifts. We have introduced new methods to estimate the bias produced by photon noise and to assess the uncertainties in the results. Our analysis suggests that, while these morphological characterisations are not precision measures
for individual clusters, they can still provide useful statistics to study trends of properties in samples of clusters. This is evident from both the results of the Monte Carlo Poisson noise simulations to estimate uncertainties, and from the difference in the substructure measures of simulated clusters when  obtained from different viewing angles.

In a comparison of the morphological parameters $P_3/P_0$ and $w$, we find typical uncertainties 
(for good {\it XMM-Newton} data quality with high 
photon statistics) of $\sim 70$ per cent and $ \sim 15$ per cent, for power ratios and $w$, respectively.
Testing the recovery of a given substructure measure from the observation of a (simulated) 
cluster from different viewing angles, we find that the difference for the different viewing 
angles is about as large as the values of $P_3/P_0$ itself, while it is smaller by about a 
factor of two for $w$. Yang et al. (2008) have recently found similar results using power 
ratios and centroid shift tests on simulated clusters with known merger histories. They 
find a substantial and significant correlation of $P_2/P_0$, $P_3/P_0$, and $w$ with the 
time passed since the last major merger (for a mass ratio smaller than 5:1). Similar to 
our findings the correlation is not tight enough for a cluster by cluster identification 
of the dynamical state, but it provides important statistical diagnostics. In addition 
they find that $w$ is significantly more sensitive than the power ratios. For our work 
with the \rexcess\ sample we have therefore adopted a threshold value of $w > 0.01$ for 
the designation of a galaxy cluster as being dynamically disturbed.

Despite the fact that $w$ is a more sensitive substructure diagnostic, one should not easily 
conclude that power ratios should be given up as an alternative. For multi-peaked (simulated) 
clusters, the power ratios pick up the obvious substructure with higher sensitivity than the 
centroid shifts, as can be seen in the larger relative excesses of the $P_3/P_0$ values compared 
to the $w$ values in Figs.~7 and~8, respectively. For the substructure - cool core correlation 
$w$ shows a much stronger connection (likely because it is less biased toward the signal from 
the outer radii), but it is $P_3/P_0$ that is more affected by the correlation with X-ray luminosity (Fig. 12). 
This illustrates that the different methods of characterising substructure react differently to 
various morphologies and it may still be useful to look at several substructure tests in cluster 
morphology studies. 

Our results also point towards a possible future improvement in our application of the power ratio method. Firstly, as shown in the Appendix, running the power ratio analysis in several apertures tends to emphasise different structural features. Furthermore, we have several cases of highly disturbed clusters in our sample, where we see large quadrupole and/or octopole moments but no significant hexapole signal, because the distortion preserves some mirror symmetry.
This suggests that it might be worthwhile to investigate a composite power ratio measure that combines the several multipole moments at various radii. This idea
can be seen in analogy to the definition of the centre shift parameter, which is also derived from statistics of measurements with several apertures.   

Finally we reemphasise that the clear statistical difference between the cool core versus non-cool core clusters, plus that of observed versus simulated clusters, provides a nice illustration of the power of the substructure measures as a statistical diagnostic of the sample under consideration. 

\subsection{Relations between cluster properties}

One of the most interesting findings of this study is that there is no
mass dependence of the substructure statistics. This is not
only revealed by the observed cluster sample, which might still be affected by
small number statistics, but is also shown by the simulated cluster sample, as shown in Fig. 15. 
Naively one might expect in the standard cosmological model of hierarchical structure
formation, where the largest structure are the youngest, that larger clusters
have had more recent mergers and therefore show on average larger substructure 
measures. However, some theoretical studies show that this might in fact be a very mild 
effect. Guo (2009) has for example studied the growth of dark matter halos via
major mergers (defined by mass ratios less than three) in the Millenium simulations (Springel et al. 2005), and finds that the merger rate differs by less than  $20$ per cent for mass differences of a factor of four in the mass range relevant for our sample. This indicates that our finding may be well consistent with the currently
adopted structure formation scenario.

The independence of cluster mass and morphology in our sample is then also reflected by the fact that there is no mass bin which has preferentially more cool cores.
This is slightly different from the result of an analysis of the 106 brightest known
galaxy clusters (the HIFLUCGS sample), where a bias towards more cool core clusters in low mass systems was found (Chen et al. 2007).

The very low correlation of substructure with cluster mass is good news for the application
of galaxy clusters to cosmology. One of the most critical tasks in these cosmological studies is the construction of robust relations between simple cluster observables and cluster 
mass. The most problematic issue here is that observed cluster samples 
consist of a mixture of relaxed and unrelaxed clusters. In this context it is of great help for the construction and calibration of observable - mass relations
to know that the ratio of relaxed to unrelaxed clusters is not a strong function of mass. 

In studying the correlation of cool core properties with substructure this paper adds a new dimension to previous work in two respects: (i) by using a morphologically
unbiased sample with selection purely by X-ray luminosity, and (ii) by analysing
cluster properties out to a fiducial global radius, $R_{500}$. Cool core objects unsurprisingly show a very regular appearance in low exposure images or in images obtained with earlier 
X-ray observatories, since in such exposures we mainly detect the bright centre, which is always very regular. In contrast to this, our study involves the cluster appearance out to a large radius, and we have carefully examined the effect of cool cores by providing results where the cool core has been 
excised. This provides a new look at some facets of the cool core cluster morphology relation problem. For example, in the present sample, we do indeed find globally disturbed clusters which harbour cool cores.
 
\subsection{Observations versus simulations}

The comparison of observations and simulations in Section~4.4 provides significant evidence that the physical recipes used to model the evolution of the intracluster medium by cooling, feedback and diffusive transport processes differs from the processes prevailing
in nature. Since the prime goal of the simulations was to reproduce the observed scaling relations of the various global ICM properties, cluster morphological parameters have rarely been used to tune the simulation recipes. One easily-identified difference between the simulations and the observations is the presence of more pronounced cool regions in the simulated clusters, as explained above. This was also seen in Pratt et al. (2007), where almost all simulated clusters feature central temperature drops, whereas only about a third of the observed clusters have cool cores.

As a consequence of the pronounced cool cores in the simulated objects, we find
merger remnants with two or more cool cores which are not yet or incompletely
disrupted. One of the \rexcess\ clusters excluded from the present analysis, RXCJ2152.2-1942, is a double cluster, but the two cluster centers are well
separated and lie outside each others $R_{500}$, which distinguishes this cluster from the simulation examples shown in Fig.~9. Similarly, the other excluded cluster RXCJ0956.4-1004 (the A901/902 supercluster) shows three well-separated X-ray emission regions. 

The simulations used here were performed in 2004. Since then new recipes have
been introduced to cosmological cluster simulations, including a significant feedback from central AGN to partly suppress the formation of cool cores. This has recently been explored both semi-analytically (e.g. Croton et al. 2006, Bower et al. 2006), and also in N-body/hydrodynamical simulations (e.g. Sijacki et al. 2008, Puchwein et al. 2009, Fabjan et al. 2009).
It will be one of our next projects to extend the comparison to this new generation of simulations. 

The lower degree of substructure seen in observations, as compared to simulations is advantageous for the cosmological application of galaxy clusters. Some of the problems pointed out in simulation studies using numerical prescriptions similar to the simulations used here, such as e.g. the effect of multi-temperature structure on the determination of cluster masses from X-ray observations (Mazzotta et al. 2004, Rasia et al. 2005, 2006) should be less problematic if extra-central cool core regions in clusters are rare in nature, and the temperature distribution in clusters is in general more regular.
This highlights the importance of testing the compatibility of the simulated clusters with representative samples of observed objects using all possible observational characteristics, to ensure their physical similarity.

\section{Conclusions}

We have used power ratios and centre shifts to investigate the substructure and 
morphological characteristics of 31 clusters from the \rexcess\ galaxy cluster sample. 
We examine in parallel a sample of 117 clusters identified from hydrodynamical 
simulations of a $\Lambda$CDM model. Substructure measures are estimated consistently 
within a radius of $R_{500}$. Our main conclusions are as follows:

\begin{itemize}

\item Using a newly-developed Monte Carlo procedure to estimate the uncertainties, we find 
that $w$ is more sensitive than power ratios for the good quality cluster images we have at 
our disposal, although combination of the two methods gives complementary information. 

\item Neither substructure measure gives an exact quantification of a cluster's dynamical 
state, and so they should only be used in a statistical sense.

\item For both observed and simulated cluster samples, the substructure parameters do not 
exhibit a mass dependence, a result that has important implications for the construction 
and calibration of the observable-mass relations for use in cosmological applications.

\item Cool core objects are generally the most regular. However there exist cool core 
systems that are identified as disturbed using both $w$ and power ratio substructure statistics.

\item As compared to the observations, the simulations contain many more cool, dense regions. 
This contributes to a statistical enhancement in the amount of substructure in the simulated 
clusters as compared to the observed objects, indicating that numerical prescriptions do not 
precisely reproduce the structure of the real cluster population.

\end{itemize}

Finally, we re-emphasise that in the present work we could only obtain statistically meaningful 
results because (i) we deal with a statistically representative sample, (ii) we have good photon 
statistics from deep {\it XMM-Newton} observations of relatively bright, not too distant clusters, 
and (iii) the data quality of the observed sample is fairly homogeneous. Any deviation from these 
ideal conditions would have made the analysis more difficult and less reliable.

\begin{acknowledgements} 

The paper is based on observations obtained with {\it XMM-Newton}, an ESA 
science mission with instruments and contributions directly funded by 
ESA Member States and the USA (NASA). The {\it XMM-Newton} project is 
supported in Germany by the Bundesministerium f\"ur Bildung und Forschung, 
Deutsches Zentrum f\"ur Luft und Raumfahrt (BMBF/DLR), the Max-Planck 
Society and the Haidenhain-Stiftung. 
GWP acknowledges partial  support from DfG Transregio Programme TR33. HB acknowledges  support for the research group through The Cluster of Excellence `Origin  and Structure of the Universe', funded by the Excellence Initiative of 
the Federal Government of Germany, EXC project number 153. K.D. acknowledges
support from the DfG Priority Programm SPP 1177.
 
\end{acknowledgements}


\appendix
\section{Further details of the data analysis}

In addition to Table~1 in the main text of the paper, we provide here the complementary Table A.1, where the substructure parameters are listed for the analysis where central regions were not excluded. For the power ratios the 
differences are in almost all cases not larger than the uncertainties. For the
$w$ parameter the results are also similar, but due to the fact that this parameter is less biased towards large radii and the better precision with which
this parameter can be determined, the differences are in some cases larger than the uncertainties. 

\begin{table*}
\begin{minipage}{\textwidth}
\begin{center}
\caption{Substructure parameters for 31 clusters from the \rexcess\ sample. The cluster centers were not excised for the analysis results shown in this table.}
\label{TabR1e_1}
\centering
\begin{tabular}{l r r r r r r r r r r r }
\hline
\hline
\\
\multicolumn{1}{l}{Cluster} & 
\multicolumn{1}{c}{$P2/P0$} & 
\multicolumn{1}{c}{bias} & \multicolumn{1}{c}{error} & 
\multicolumn{1}{c}{$P3/P0$} & 
\multicolumn{1}{c}{bias} & \multicolumn{1}{c}{error} & 
\multicolumn{1}{c}{$P4/P0$} & 
\multicolumn{1}{c}{bias} & \multicolumn{1}{c}{error} &
\multicolumn{1}{c}{$w$} & \multicolumn{1}{c}{error} \\

\multicolumn{1}{c}{} & 
\multicolumn{1}{c}{$\times 10^{-5}$} & 
\multicolumn{1}{c}{$\times 10^{-5}$} & \multicolumn{1}{c}{$\times 10^{-5}$} & 
\multicolumn{1}{c}{$\times 10^{-7}$} & 
\multicolumn{1}{c}{$\times 10^{-7}$} & \multicolumn{1}{c}{$\times 10^{-7}$} & 
\multicolumn{1}{c}{$\times 10^{-7}$} & 
\multicolumn{1}{c}{$\times 10^{-7}$} & \multicolumn{1}{c}{$\times 10^{-7}$} &
\multicolumn{1}{c}{} & \multicolumn{1}{c}{} \\

(1) & (2) & (3) & (4) & (5) & (6) & (7) & (8) & (9) & (10) & (11) & (12)\\
\\
\hline
\\
{\rm RXCJ0003.8+0203} &   0.131 &   0.0041 &   0.031 &   0.220 &   0.1318 &   0.265 &   0.3395 &   0.058 &   0.213 &   0.0028 &  0.00079\\
{\rm RXCJ0006.0-3443} &   0.448 &   0.0072 &   0.075 &   1.862 &   0.2348 &   1.055 &   0.2179 &   0.101 &   0.222 &   0.0190 &  0.00153\\
{\rm RXCJ0020.7-2542} &   0.093 &   0.0033 &   0.025 &  -0.094 &   0.1152 &   0.198 &   0.2389 &   0.050 &   0.215 &   0.0168 &  0.00116\\
{\rm RXCJ0049.4-2931} &   0.091 &   0.0068 &   0.040 &   0.115 &   0.2156 &   0.477 &   1.0930 &   0.093 &   0.533 &   0.0026 &  0.00067\\
{\rm RXCJ0145.0-5300} &   1.054 &   0.0057 &   0.113 &   1.000 &   0.1748 &   0.689 &   1.6230 &   0.078 &   0.556 &   0.0297 &  0.00170\\
{\rm RXCJ0211.4-4017} &   0.307 &   0.0054 &   0.063 &  -0.142 &   0.1871 &   0.213 &   1.3600 &   0.083 &   0.480 &   0.0041 &  0.00078\\
{\rm RXCJ0225.1-2928} &   0.588 &   0.0146 &   0.121 &   4.883 &   0.4361 &   2.374 &   1.3040 &   0.239 &   0.759 &   0.0114 &  0.00139\\
{\rm RXCJ0345.7-4112} &   0.152 &   0.0046 &   0.033 &   1.403 &   0.1572 &   0.743 &   0.8983 &   0.074 &   0.405 &   0.0044 &  0.00063\\
{\rm RXCJ0547.6-3152} &   0.081 &   0.0019 &   0.016 &   1.243 &   0.0543 &   0.348 &   0.7068 &   0.023 &   0.169 &   0.0129 &  0.00081\\
{\rm RXCJ0605.8-3518} &   0.064 &   0.0011 &   0.010 &   0.007 &   0.0298 &   0.056 &   0.0271 &   0.014 &   0.029 &   0.0057 &  0.00022\\
{\rm RXCJ0616.8-4748} &   0.372 &   0.0063 &   0.084 &   6.137 &   0.1969 &   1.697 &   2.9580 &   0.088 &   0.742 &   0.0161 &  0.00143\\
{\rm RXCJ0645.4-5413} &   0.258 &   0.0024 &   0.034 &  -0.068 &   0.0720 &   0.095 &   0.1835 &   0.029 &   0.106 &   0.0121 &  0.00049\\
{\rm RXCJ0821.8+0112} &   0.153 &   0.0188 &   0.083 &   6.717 &   0.5968 &   3.133 &   1.8220 &   0.303 &   1.075 &   0.0062 &  0.01416\\
{\rm RXCJ0958.3-1103} &   0.100 &   0.0031 &   0.022 &   0.014 &   0.0922 &   0.195 &   0.1165 &   0.042 &   0.101 &   0.0029 &  0.00054\\
{\rm RXCJ1044.5-0704} &   0.116 &   0.0008 &   0.011 &  -0.013 &   0.0216 &   0.019 &  -0.0029 &   0.009 &   0.019 &   0.0042 &  0.00024\\
{\rm RXCJ1141.4-1216} &   0.041 &   0.0013 &   0.010 &   0.500 &   0.0399 &   0.198 &   0.0535 &   0.016 &   0.049 &   0.0052 &  0.00051\\
{\rm RXCJ1236.7-3354} &   0.025 &   0.0051 &   0.018 &   0.145 &   0.1825 &   0.374 &   0.1157 &   0.081 &   0.244 &   0.0048 &  0.06762\\
{\rm RXCJ1302.8-0230} &   0.722 &   0.0042 &   0.080 &   1.986 &   0.1345 &   0.647 &   0.8242 &   0.062 &   0.341 &   0.0215 &  0.02096\\
{\rm RXCJ1311.4-0120} &   0.017 &   0.0004 &   0.003 &   0.024 &   0.0096 &   0.025 &   0.0066 &   0.004 &   0.010 &   0.0029 &  0.00026\\
{\rm RXCJ1516.3+0005} &   0.075 &   0.0025 &   0.018 &   0.315 &   0.0710 &   0.235 &   0.2417 &   0.033 &   0.132 &   0.0059 &  0.00054\\
{\rm RXCJ1516.5-0056} &   0.534 &   0.0065 &   0.076 &   8.604 &   0.2268 &   1.691 &   0.8404 &   0.091 &   0.394 &   0.0160 &  0.00135\\
{\rm RXCJ2014.8-2430} &   0.026 &   0.0007 &   0.006 &   0.206 &   0.0203 &   0.091 &   0.0427 &   0.009 &   0.034 &   0.0053 &  0.00022\\
{\rm RXCJ2023.0-2056} &   0.042 &   0.0118 &   0.035 &  -0.199 &   0.4147 &   0.448 &   0.5655 &   0.178 &   0.594 &   0.0191 &  0.00130\\
{\rm RXCJ2048.1-1750} &   0.760 &   0.0050 &   0.073 &   3.350 &   0.1311 &   0.813 &   1.8790 &   0.055 &   0.429 &   0.0460 &  0.00494\\
{\rm RXCJ2129.8-5048} &   0.034 &   0.0059 &   0.052 &   2.631 &   0.2068 &   1.115 &  -0.0427 &   0.096 &   0.160 &   0.0479 &  0.02450\\
{\rm RXCJ2149.1-3041} &   0.005 &   0.0019 &   0.005 &   0.793 &   0.0538 &   0.262 &   0.0218 &   0.024 &   0.038 &   0.0038 &  0.00047\\
{\rm RXCJ2157.4-0747} &   1.376 &   0.0240 &   0.251 &  -0.691 &   0.7871 &   0.862 &   3.1270 &   0.392 &   1.709 &   0.0517 &  0.00343\\
{\rm RXCJ2217.7-3543} &   0.039 &   0.0025 &   0.014 &   0.449 &   0.0658 &   0.237 &   0.1529 &   0.029 &   0.122 &   0.0019 &  0.00047\\
{\rm RXCJ2218.6-3853} &   0.492 &   0.0016 &   0.035 &   0.351 &   0.0439 &   0.189 &   0.1159 &   0.021 &   0.063 &   0.0188 &  0.00057\\
{\rm RXCJ2234.5-3744} &   0.101 &   0.0030 &   0.018 &  -0.060 &   0.0900 &   0.129 &   0.2564 &   0.040 &   0.102 &   0.0136 &  0.00150\\
{\rm RXCJ2319.6-7313} &   0.340 &   0.0047 &   0.058 &  -0.119 &   0.1469 &   0.154 &   0.2151 &   0.065 &   0.178 &   0.0187 &  0.00103\\

\\
\hline
\end{tabular}
\end{center}

NOTES: The power ratio parameters have been determined for an aperture with a radius of $R_{500}$. The corresponding results without center excision are given in Table~\ref{TabR1e}. For each of the power ratio parameters we provide the value of the noise contribution to the power ratio result (bias) which has been subtracted from the measured result to provide the value listed in columns 2, 5, and 8. The uncertainties determined from the Poissonisation simulations are listed in columns 4, 7, and 10 (error). The center shift statistic parameter $w$ and its uncertainty are listed in columns 11 and 12.

\end{minipage}
\end{table*}

\subsection{Bias and error estimation for the power ratio method\label{appx:a1}}

   \begin{figure}
   \begin{center}
   \includegraphics[width=\columnwidth]{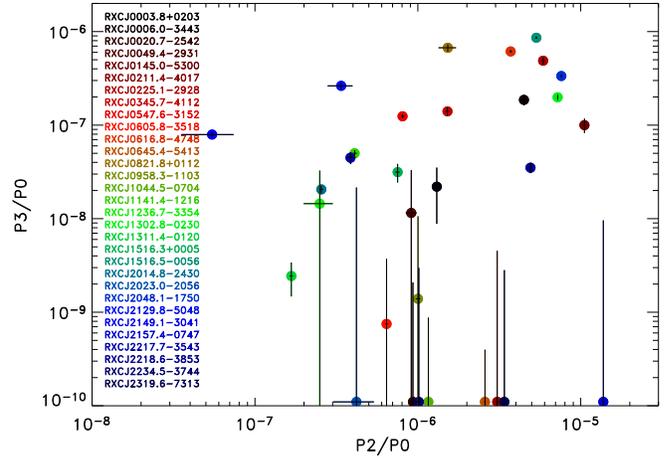}
      \caption{$P_2/P_0$ and $P_3/P_0$ for the 31 galaxy clusters of the \rexcess\ sample. The errors shown are those from simulations with azimuthal randomization used to test against the null hypothesis that the cluster is azimuthally symmetric. After correction for  the signal contribution from photon noise, 8 clusters in the sample have negative values of $P3/P0$ that are consistent with zero within the error limits.
       These data points are indicated at the bottom of the plot.}
         \label{FigA1}
  \end{center}
   \end{figure}

In this Appendix, we explain the new methods we have used to estimate photon noise bias and substructure measure uncertainties.
Having performed the power ratio analysis as outlined in Equations~1 to~4,
the first question we should ask in interpreting the results is: when have we detected a significant signal of deviation from azimuthal symmetry? Even a completely symmetric
cluster in nature would be detected with some residual structure due to the photon noise
with which it is observed. To asses this we perform the following test. We conduct
a second substructure analysis in which for all flux pixels entering
the integrals of Eqns.~3 and~4 we take only the radius from the data but assign randomly drawn angles $\phi$. This randomises all azimuthal structure the cluster might have. We repeat this process 1\,000 times and determine the mean and dispersion of the distribution of the power ratio parameters.
The mean of the signal can be interpreted as the typical residual signal
a perfectly regular cluster would have in the presence of photon noise. We interpret this spurious signal as a
measure of the typical photon noise contribution to the power ratio measurement in all
clusters and subtract its value from the obtained result to recover the intrinsic signal of the power ratio. The standard deviation of the spurious signal
from the mean in all simulations is used as a first estimate of the uncertainty of the final net result. Some very regular clusters come out with a negative signal after
subtraction of the photon noise contribution. But this negative signal is never 
larger than the uncertainty.      

Fig.~A.1 shows $P_2/P_0$ and $P_3/P_0$ for the \rexcess\ sample,
with uncertainties estimated from the azimuthal randomisation process. The mean signal bias and associated uncertainties are always very similar, which is not surprising. Inspecting the uncertainties we note that for the data quality of the \rexcess\
sample we need typical values of $P_2/P_0 \ge 10^{-7}$, $P_3/P_0 \ge 2 - 4 \times 10^{-8}$ and $P_4/P_0 \ge 2 \times 10^{-8}$ in order to be able to claim significant detection of substructure or deviations from azimuthal symmetry. 

The above considerations provide a useful uncertainty estimate for clusters which have
substructure signals close to the significance threshold. If a cluster has a strong 
substructural feature, the uncertainty of the measurement of the substructure parameter
will depend critically on the photon noise connected to this feature, which is probably not related to the photon noise of a randomized cluster where this feature has been
washed out. Therefore we need a new approach to estimate uncertainties.
To test what uncertainties are expected for clusters with larger signals, we performed the following simulations. We used the combined count image (without background subtraction) and obtained a new integer random number for the photon counts in each pixel by drawing the numbers
from a Poissonian distribution with the  observed photon counts as expectation values.
The Poissonized image is background and exposure corrected and subjected to precisely the same
substructure analysis as the observed clusters. The process was repeated 200 times, and the mean and standard deviation determined. 
The mean simulated values are slightly higher than the observed values. This is due to the fact that the simulated images contain
the artificial Poissonized noise on top of the observational noise, which means their mean noise
added structure parameters should be slightly biased high. However, the scatter is expected to give a good representation of the Poisson noise uncertainty of the observed data. We therefore use the standard deviation of these 200 Poisson simulations as a measure of the uncertainty for the power ratio parameters. Fig.~1 shows the same data as Fig.~A.1 but with the uncertainties determined from the Poissoniation simulations. The errors are considerably larger for the clusters with highly significant signals, but are similar near the significance threshold. Therefore our suspicion was correct
that we need a different assessment for the measurement uncertainties than that given simply by the bias.

   \begin{figure}
   \begin{center}
   \includegraphics[width=\columnwidth]{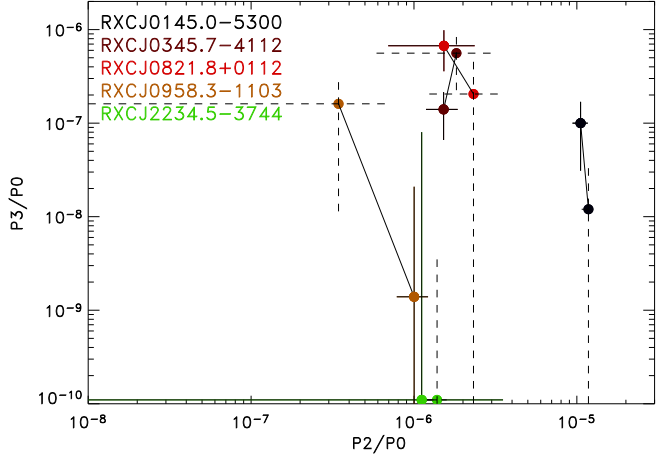}
      \caption{Comparison of the power ratio results $P2/P0$ and $P3/P0$ for
         five of the \rexcess\ clusters with two separate exposures. The results
         approximately agree within the combined errors, supporting our approach of
         estimating uncertainties with Poisson Monte Carlo simulations.}
         \label{FigA2}
  \end{center}
   \end{figure}

We can further check if the large uncertainties that we obtained are realistic.
For this we compare the results of the analysis of five \rexcess\ clusters where we have multiple observations. These five clusters nicely cover a range of different morphologies. We show the results in Fig. A.2. The results for separate observations of the same cluster
are different but all overlap with their uncertainties. The second observations
have in some cases significantly lower exposure times, which increases the uncertainty.
This test illustrates that the relatively large error estimates we calculate for the power ratios are well-justified. 

We note once again that these uncertainties come from an
end-to-end test of the analysis, since all the analysis steps of the power ratio
determination are performed on the Poisson resampled images (i.e., including the centring and bias subtraction on top of the application of the power ratio formulae).  

\subsection{Influence of the aperture radius and angular resolution on the power ratio results} 

In the next test we explored the effect that the angular resolution of the observation has on the results. Using four clusters which again cover an interesting range of morphologies we determined the power ratios as a function of increased smoothing of the cluster images. The results are displayed in Fig.~A.3. The change of the  power ratios is shown as a function of successive smoothing by a Gaussian of width 4, 8, 15, 30, and 60 arcsec. For all four clusters the change is small, smaller than or roughly comparable to 
the typical errors of the overall measurement. Therefore the angular resolution of the observations
is not an issue for the sample or for the comparison to other observations 
(e.g. clusters at higher redshifts observed at lower angular resolution).

   \begin{figure}
   \begin{center}
   \includegraphics[width=\columnwidth]{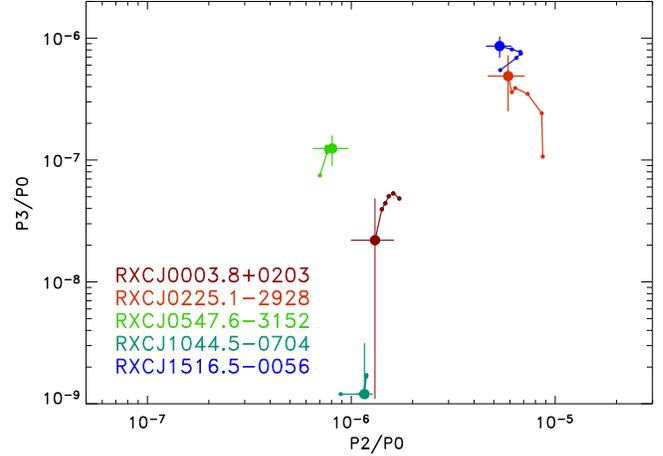}
      \caption{Influence of the sampling point spread function (PSF) of the X-ray images on the derived power ratios. The heavy points give the unsmoothed
       result with {\it XMM-Newton} resolution. The smaller points connected by a line show the effect of successive smoothing by a Gaussian of width 4, 8, 15, 30, 60 arcsec width. The error bars
       shown for the unsmoothed data points are those obtained from the Poissonisation simulations.}
         \label{FigA3}
  \end{center}
   \end{figure}

   \begin{figure}
   \begin{center}
   \includegraphics[width=\columnwidth]{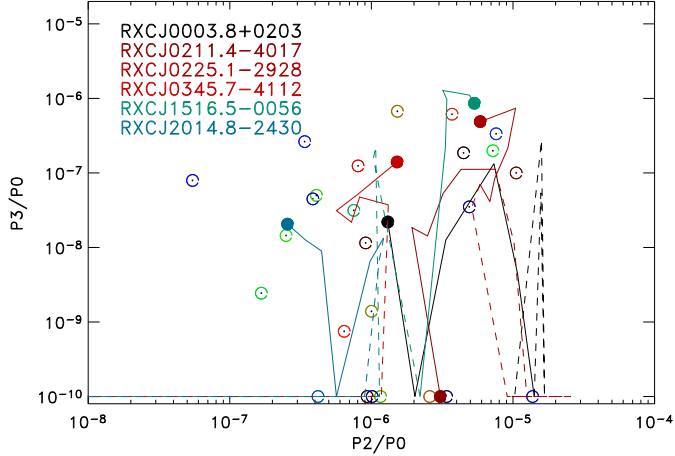}
      \caption{Effects of changing the aperture radius, $R_{\rm ap}$, on the power ratios illustrated for six of the \rexcess\ clusters with different morphologies. The aperture
      radius is decreased from $R_{500}$ (marked as solid dot) to $0.5 R_{500}$ in steps of $0.1 R_{500}$ (shown as solid line) and further decreased to  $0.1 R_{500}$ (shown as dashed line). The parameter distribution for the remaining \rexcess\ clusters is indicated by open symbols.}
         \label{FigA4}
  \end{center}
   \end{figure}

To obtain an overview on the dependence of the power ratios on the aperture radius, we have calculated the power ratios for 10 different radii for all clusters starting with $0.1\times R_{500}$ and increasing in
steps of $0.1\times R_{500}$. Due to the large powers of $R^{\prime}$ that appear in
Eqs.~3 and~4, structure near the aperture radius is most heavily weighted. This
became very obvious in the first tests we performed before removing point sources.  Even only moderately strong
point sources near $R_{\rm ap}$ have a clear effect on the orientation of the multipoles and this effect decreases for smaller radii. Therefore we can expect that different 
structural features in the clusters become important for different values of $R_{\rm ap}$.
We have clearly seen this in the visual inspection of the results. Fig.~A.4 shows the change of the power ratios with $R_{\rm ap}$ for some examples of the \rexcess\ sample.
Since we are mostly interested in global cluster parameters, e g. in the study of the correlation
of the structure parameters with other global parameters determined within $R_{500}$ (e.g. Pratt et al. 2007, 2009a), we have concentrated on the results obtained
for $R_{\rm ap} = R_{500}$. Small changes in $R_{\rm ap}$ leave the clusters in the same parameter
range, while changes of the order of $0.5 R_{500}$ can give quite different results.

\subsection{Comparison of power ratios and centre shifts}
 
In comparing the results of the power ratios with the centre shifts, we were anticipating that centre shift measures would be more sensitive to the central
regions while power ratios are most sensitive to the outermost zones. To investigate this in more detail, we have looked at the correlation of $P_3/P_0$  with $w$, as shown in Fig.~2 for an aperture of $R_{500}$, but now as a function of the aperture radius. Table~3 gives the correlation coefficients for six aperture radii between $0.5\,R_{500}$ and $R_{500}$. We clearly note a very sharp maximum of the correlation coefficient for an aperture radius between 0.7 - $0.8\,R_{500}$. The distribution of the two substructure measures for an aperture radius $R_{\rm ap} = 0.7\,R_{500}$ is shown in Fig.~A.5. The points are now visually more correlated than in the corresponding Fig.~2. We also note a that cool cores are more clearly as being less morphologically disturbed; the classification of the power ratio parameter is now practically as tight as that of the $w$ parameter.  The only cool core cluster that is classified by both methods as disturbed is RXCJ1302.8-0230, which appears in the upper right quandrant.This system has a cool core that is clearly off-set from the large scale cluster center.
These results thus seem to further encourage 
us to think about useful combinations of power ratio parameters from different 
radii to construct a more sensitive substructure measure.    

   \begin{figure}
   \begin{center}
   \includegraphics[width=\columnwidth]{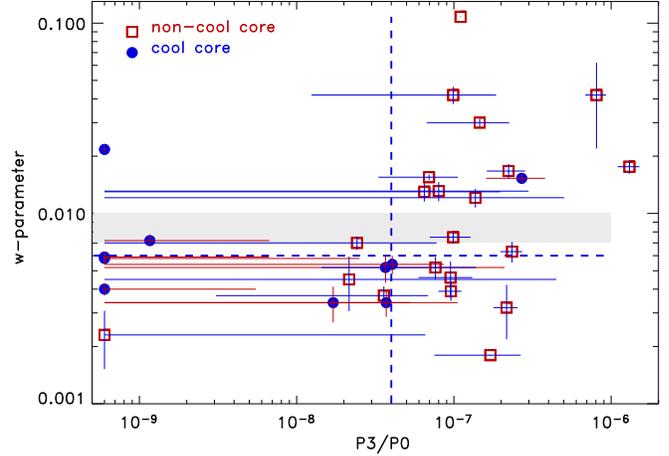}
      \caption{$w$ versus $P_3/P_0$
    for the 31 clusters of the \rexcess\ sample. This Figure is similar to Fig.~2, but now $P_3/P_0$ is determined for an aperture radius $R_{\rm ap} = 0.7\,R_{500}$. The dashed blue lines and the grey bar are the same as in Fig.~2 for better comparison. The parameter values shown here were derived for the core excised images. We identify the clusters by their cool core properties
     as explained in the text. 
}
         \label{FigA5}
  \end{center}
   \end{figure}

\end{document}